\documentclass{aa}

\usepackage{txfonts}
\usepackage{natbib}
\bibpunct{(}{)}{;}{a}{}{,} 
\usepackage{graphicx}

\usepackage{soul}

\newcommand{\ie}{i.e.~\/}
\newcommand{\kms}{${\rm km~s^{-1}}$}
\newcommand{\eg}{e.g.,}
\newcommand{\etal}{{\em et al.}}
\newcommand{\poh}{\phantom{0}}

\begin{document}

\title{Variability in X-ray line ratios in helium-like ions 
of massive stars: the radiation-driven case}

\titlerunning{Radiation-driven X-ray line ratio variability}
\authorrunning{K. T. Hole \& R. Ignace}

%

\author{K. T. Hole\inst{1}
\and
R. Ignace\inst{1}}

\institute{Department of Physics and Astronomy, East Tennessee State University
	P. O. Box 70652, Johnson City, TN 37604, USA}

\offprints{K. T. Hole, \email{holekt@etsu.edu}}

\date{Received <date> / 
Accepted <date>}

 \abstract
   {
Line ratios in ``$fir$'' triplets of helium-like ions have proven to
be a powerful diagnostic of conditions in X-ray emitting gas
surrounding massive stars. Recent observations indicate that these
ratios can be variable with time.
   }
   {
The possibilities for causes of variation in line $ratios$ are
limited: changes in the radiation field or changes in density, which
would result in radiational or collisional level-pumping,
respectively; and changes in mass-loss or geometry, which could cause
variations in X-ray absorption effects.  In this paper, we explore the
first of these potential causes of variability: changes in the
radiation field. We therefore explore the conditions necessary to
induce variability in the ratio $R=f/i$ via this mechanism.
   }
   {
To isolate the radiative effect, we use a heuristic model of
temperature and radius changes in variable stars in the B and O range
with low-density, steady-state winds. We then model the changes in
emissivity of X-ray emitting gas close to the star due to differences
in level-pumping owing to the availability of UV photons at the
location of the gas.
   }
   {
We find that under these conditions, variability in $R$ is
dominated by the stellar temperature.  Although the relative amplitude
of variability is roughly comparable for most lines at most
temperatures, detectable variations are limited to a few lines for
each spectral type.  We predict that variable values in $R$ due to
stellar variability must follow predictable trends found in our
simulations.
   }
   {
Our heuristic model uses radial pulsations as a mode of stellar
variability that maximizes the amplitude of variation in $R$.  This
model is robust enough to provide a guide to which ions will provide
the best opportunity for observing variability in the $f/i$ ratio at
different stellar temperatures, and the correlation of that
variability with other observable parameters.  In real systems,
however, the effects would be more complex than in our model, with
differences in phase and suppressed amplitude in the presence of
non-radial pulsations.  This, combined with the range of amplitudes
produced in our simulations, suggest that changes in $R$ across many
lines concurrently are not likely to be produced by a variable
radiation field.
   }

\keywords{Line: profiles; Radiation mechanisms: general; 
Stars: early-type; X-rays: stars; Atomic processes; 
Line: formation 
}

\maketitle

%
\section{Introduction}
\label{s:introduction}

Astrophysical studies at X-ray wavelengths have been revolutionized
through high resolution spectroscopy afforded by the {\em Chandra} and
{\em XMM-Newton} telescopes.  Diagnostics for temperature, density,
flow dynamics, and geometry can be observed with these telescopes
through the use of resolved spectral lines from X-ray emitting
plasmas.  Our understanding of massive star winds in particular has
been impacted by these new observational capabilities.

With the large investment of observing time in existing studies of hot
plasma emissions from massive stars, we believe that the next natural
step in understanding these X-ray sources is to consider their
spectral variability.  This is in many ways a fledgling field, since
substantial amounts of observing time is required merely to resolve
lines.  Still, some sources such as $\zeta$~Pup have been observed in
excess of 1~Msec over a period of many years.  The fast winds of
massive stars have a flow time given by the ratio of the stellar
radius $r_\star$ to the wind terminal speed $v_\infty$, with
$r_\star/v_\infty \sim$ ksec timescales.  Many of these stars also
have rotation speeds of hundreds of \kms, implying rotational periods
of days to weeks, on order of a Msec.  Thus there is good physical
motivation to investigate X-ray variability for these sources at these
timescales.

There have already been several observational studies of X-ray
variability among massive star sources.  Some examples include
colliding wind binaries (e.g., $\eta$~Car: \citealt{henley2008},
\citealt{Corcoran2001}, \citealt{leutenegger2003}; $\gamma$~Vel:
\citealt{skinner2001}, \citealt{henley2005}, \citealt{schild2004},
\citealt{stevens1996}; WR~140: \citealt{pollock2005}; WR~147:
\citealt{zhekov2010}, \citealt{skinner2007}), mass transfer binaries
(e.g., $\beta$~Lyr: \citealt{ignace2008}), magnetic stars (e.g.,
$\theta^1$ Ori C: \citealt{gagne2005, ignace2010}; $\sigma$~Ori E:
\citealt{groote2004}; $\tau$~Sco: \citealt{ignace2010},
\citealt{mewe2003b}, \citealt{cohen2003}), or pulsator stars (e.g.,
\citealt{favata2009}; Oskinova \etal\ in press).  These sources tend
to be unusual in some manner such that cyclical variability is to be
expected.  Some variability studies have also targeted nominally
single stars (e.g., $\zeta$~Pup: \citealt{berghoefer1996}; WR~46:
\citealt{gosset2011}).

Although not true in every case, many of these variability studies
have been limited to considerations of X-ray passband variability or
variability of low resolution spectral energy distributions (SEDs).
There have been some exceptions, however, notably the work of
\citet{henley2008} on $\eta$~Car or the \citet{favata2009} data set on
$\beta$~Cep, in which variations are seen in some individual emission
lines.

New reports of spectral line variability are emerging, such as that by
\citet{nichols2011}, who analyzed archival data for a number of
massive star sources, and a study of $\kappa$~Ori by Waldron \etal\
(priv.\ comm.).  Although the data are not yet adequate to conduct a
full period analysis, it does appear that the $R$ ratio (discussed
below) has detectable variations in several helium-like ions at the
same time from putatively-single massive stars.  As a ratio involving
components from a line triplet, such variations cannot simply arise as
a result of emission measure (EM) variations. Instead, they are
sensitive to other important effects that provide fresh diagnostic
value for understanding massive star X-ray emission processes.  While
these reports of line variability are still undergoing analysis, it
does seem timely to engage in a discussion of the {\em theoretical}
expectations for line variability.

While there are many possibilities for the causes of such variability,
in this first contribution we limit the scope of our discussion to how
the triplet lines of He-like species can vary in response to a
changing radiation field.  These X-ray lines, which include the
forbidden, intercombination, and resonance (``$fir$'') triplets of
abundant metal species such as \ion{N}{vi}, \ion{O}{vii}, \ion{Ne}{ix}
and so on, are typically quite strong. There is a fairly broad range
of X-ray temperatures in which these species can be the dominant stage
of ionization \citep[\eg][]{cox2000}.  Thus the triplet lines have
long been used as important diagnostics of the hot plasma.

\citet{gabriel1969} discussed two key line ratios that could be formed
from the triplet.  The first was a ratio of the forbidden line
emission to that of the intercombination line, with $R=f/i$.  The
second involved a ratio of the sum of the $f$ and $i$ line components
to that of the resonance line $r$, with $G=(f+i)/r$.  Under the
condition of collisionally dominated excitation, the ratio $G$ is
sensitive to plasma temperature, whereas the ratio $R$ is sensitive to
pumping processes that can depopulate the $f$ level to the $i$ level.
\citet{blumenthal1972} continued the work by Gabriel \& Jordan,
clarifying the concepts and emphasizing the sensitivity of $R$ to the
radiation field.

It is the radiative pumping that has proven to be of tremendous
diagnostic value in applications to massive stars
\citep[e.g.][]{kahn2001,waldron2001}.  Pumping for the most common
X-ray lines occurs in the UV, where massive stars have peak
brightnesses.  In these stars, X-ray emissions can be produced in the
wind outflow, owing to the formation of highly supersonic shocks
arising from a natural instability in the line-driving mechanism that
accounts for the wind acceleration (e.g., \citealt{lucy1982};
\citealt{owocki1988}; \citealt{feldmeier1997}; \citealt{dessart2003}).
Since the X-ray emissions are distributed above the photosphere,
radiative pumping is also a function of radius via the dilution
factor, because the pumping rate depends on the mean intensity of
radiation.  As a result, the $R$ ratio has been employed as a means of
determining or limiting the {\em location} of X-ray sources in the
circumstellar environment of massive stars.

Here we focus specifically on the $R$ ratio for a circumstellar source
of X-ray emission around a UV bright star.  There are essentially four
things that can modify line emissions in the regime of collisional
ionization equilibrium (CIE): density variations that affect the scale
of the EM; density variations that affect collisional pumping between
the levels; variations of the radiation field that affect the
radiative pumping between levels; and variations in the hot plasma
temperature.  Of these, the first two will not influence the ratios
$G$ and $R$, because variations in EM for optically thin lines is a
simple scaling factor that will cancel in the ratio.  The $R$ ratio is
sensitive to pumping effects, but $G$ is largely insensitive; on the
other hand $G$ is temperature sensitive, whereas $R$ is less sensitive
to this parameter. Thus here we are mostly concerned with variations
in the stellar radiation field at specific line pumping wavelengths.

Our goal is to develop a model that will produce the most general
results possible.  To this end, we seek to {\em maximize} the
amplitude of variability from our model.  In other words we seek to
develop a model that is as optimistic as possible for detecting
variable ratios of $R$.  We therefore consider a phenomenological
model of a B~star or O~star that is a simple radial pulsator (see
Sect. \ref{ss:sys}). The star is assumed to have a low mass-loss rate
($\dot{M}$, which allows us to ignore both photoabsorption of X-ray
emissions by the wind itself and collisional pumping effects.  This
model has the potential to affect the $R$ ratio in two ways. First,
variations in stellar temperature $T_\star$ will change the shape and
intensity of the blackbody radiation around the star. Second, changes
in stellar photospheric radius $r_\star$ can impact the dilution
factor at the location of the emitter.

A ``toy'' pulsating star model for influencing X-ray emission lines is
presented in Sect.~\ref{s:model}.  Results of our simulations are
presented in terms of a parameter study in Sect.~\ref{s:results}.  A
discussion of applications to massive stars follows in
Sect.~\ref{s:discussion}, including a qualitative discussion of
expectations when assumptions of the current model are relaxed.

\section{Model}
\label{s:model}
%

\subsection{Radiative equations}
\label{ss:eqns}

The total line emission from the triplet components are described in
terms of luminosities, $L_{f}$, $L_{i}$, and $L_{r}$ 
(or $f$, $i$, and $r$) representing the line luminosities, respectively.
The $R$ ratio is then:
\begin{equation}
\label{e:R1}
R=\frac{L_f}{L_i}.
\end{equation}
\citet{blumenthal1972} showed that in a hot plasma 
this ratio goes as
\begin{equation}
\label{e:R2}
R=\frac{R_0}{1+\phi_\nu/\phi_{\rm c} + n_{\rm e}/n_{\rm c}},
\end{equation}
where $\phi_\nu$ is the intensity of radiation at the frequency
corresponding to the transition between the $i$ and $f$ energy levels
for a given helium-like ion; $n_{\rm e}$ is the electron density; and
$R_0$, $\phi_{\rm c}$ and $n_{\rm c}$ are determined by atomic
parameters and the electron temperature of the emitting gas.  The
value of $R_0$ is a weak function of temperature under the conditions
of CIE \citep{blumenthal1972}, which are generally thought to hold for
massive star X-ray emissions.  Atomic constants for a number of common
ions used in our simulations are listed in Table \ref{t:lines}.

\begin{table*}
\begin{center}
\caption{Physical constants for six He-like ions used in our simulations.
\label{t:lines}}
\begin{tabular}{ccccccc}
\hline
Line & $\lambda_{\rm f}$ $^a$ & $\lambda_{\rm i}$ $^a$ &
 $\lambda_{\rm r}$ $^a$ 
& $A_{\rm if}$ $^b$ & $\phi_{\rm c}$ $^b$ & $F$ $^b$ \\
 & (\AA) & (\AA) & (\AA) & (s$^{-1}$) &  (s$^{-1}$) & \\
\hline\hline
\ion{C}{v} &41.46 & 40.71 &40.28  & 1.14$\times10^8$& 3.57$\times10^1$& 0.36 \\
\ion{N}{vi} & 29.53& 29.08& 28.79 & 1.37$\times10^8$& 1.83$\times10^2$& 0.38\\
\ion{O}{vii} & 22.10&21.80 &21.60 & 1.61$\times10^8$& 7.32$\times10^2$& 0.42\\
\ion{Ne}{ix} &13.70 & 13.55& 13.45& 2.12$\times10^8$& 7.73$\times10^3$& 0.41\\
\ion{Mg}{xi} &  9.31& 9.23& 9.17& 2.68$\times10^8$& 4.86$\times10^4$& 0.49\\
\ion{Si}{xiii} & 6.74& 6.69& 6.65& 3.32$\times10^8$ & 2.39$\times10^5$ & 0.49\\
\hline
\end{tabular}
\end{center}
\tablebib{
(a)~\citet{porquet2001}; (b) \citet{blumenthal1972}.
}
\end{table*}

Since our intention is to investigate how a change in the stellar
radiation field affects $R$, in our simulations we may safely assume
$n_{\rm e}\ll n_{\rm c}$ for B~star winds, as discussed in the
previous section.  The radiative pumping rate $\phi$ (in units of
photons s$^{-1}$) is defined as
\begin{equation}
\label{e:phi}
\phi= A_{tot} W\left[e^\beta-1\right]^{-1},
\end{equation}
where $A_{tot}$ is the sum of the three Einstein A-coefficients for
the transition between the $i$ and $f$ energy levels, $\beta=h\nu/k_B
T_\star$ (with $T_\star$ the time-dependent stellar effective
temperature producing a blackbody radiation field), and $W(r)$ is the
geometrical dilution factor for the intensity of the stellar radiation
field at the location of the gas.  In our simulations, we follow
Blumenthal et al. in using $A_{tot}= 3 A_{\rm if}$, with values of
$A_{\rm if}$ listed in Table \ref{t:lines}.  For a star without limb
darkening, and ignoring any diffuse radiation field in the wind
itself, the dilution factor is a simple function of radius given by
\begin{equation}
\label{e:W}
W=\frac{1}{2}\left[1-\sqrt{1- \left(\frac{r_\star}{r}\right)^2}\right]
\end{equation}

One should bear in mind some of the limitations implied by our
restrictive model.  For example, a hot gas component arising from a
shock in a wind, or from confined flow in a magnetosphere, will have a
range of temperatures.  Having a range of temperatures would create a
distribution of $R_0$ values within the gas, as well as changing the
ionization fraction.  If $R_0$ varies spatially within the gas, the
differing values of $R$ will tend to dampen the variation of $R$ with
the radiation field, making it harder to detect.  If ionization is
allowed to change with stellar temperature, that adds another
potentially confounding variable that would making it more difficult
to observe the variation.

The above-mentioned effects could be considered; however, their
inclusion would require us to choose a particular dependence of the
temperature of the X-ray emitting gas ($T_X$)with radius, making the
results model dependent.  In these simulations we omit such
considerations and follow \citet{blumenthal1972} in using the $R_0$ of
each ion at the temperature where it will have the maximum signal.
This would seem to imply physically not only that we have a
multi-temperature gas, but one in which each element had a different
temperature.  However, since the emissivity of each ion is a fairly
sharp function of temperature, while $R_0$ is a {\em weak} function of
$T$, this is a reasonable approximation of the response of the gas.
Here it is worth stressing again that there are many ways in which
variability in $R$ may be suppressed, and that our simple model is an
attempt to find the most optimistic approach for how the stellar
radiation field could drive an observable variation in the value of
$R$.  We address expansions of our results to more generalized cases
in Sect.~\ref{s:discussion}.

\subsection{System model}
\label{ss:sys}

We have chosen to use B~stars as the basis for our models. B~stars
have sufficient UV fields to cause radiative pumping, are hot enough
to produce the necessary X-ray emission, yet have low enough mass-loss
that we can avoid complications of density-induced collisional level
pumping as well as photoabsorption of X-rays within the wind. We also
extend our models to higher, O~star temperatures with the constraint
that their mass-loss remains low. This should be kept in mind when
interpreting our results for early-type stars.

Although variable early-type stars (e.g., $\beta$ Cepheids) are
non-radial pulsators, our ``toy'' radial pulsator has the advantage of
maximizing the amplitude (and thus observability) of variation in $R$.
This is because in radial pulsations, the entire surface of the star
changes in brightness, temperature and position in the same sense (\ie
brighter or dimmer) at the same time, whereas a non-radial pulsator
has some parts that are changing in opposing ways at the same time.
Thus we model radial pulsations in our stars so that our calculations
will produce {\em upper} limits to variations of $R$.

We consider a range of stellar parameters appropriate for B and
low-$\dot{M}$ O~stars: radii of 1.8 to 11.0 $r_{\sun}$ and effective
surface temperatures in the range 10,000 to 40,000~K.  For simplicity,
the radiation field is assumed to follow a blackbody curve.  Radius
and temperature variations with amplitudes of 10-20\% are considered,
typical of radially-pulsating classical Cepheid stars
\citep[\eg][]{bono2000}.  We adopt simple sinusoidal forms for the
radius and temprature evolution in angular pulsational phase $\xi$,
with:
\begin{eqnarray}
\label{e:RTvari}
T_{\star}(t)=T_{0}\left[1+\Delta T \cos{\left(\xi-\psi\right)}\right], \\
r_{\star}(t)=r_{\star,0}\left[1+\Delta r_{\star} \cos{\xi}\right].
\end{eqnarray}
Stellar pulsations, such as those caused by the $\kappa$-mechanism,
generally have a phase offset between changes in the radius and in the
effective temperature of the star.  Our model allows for such an
offset by varying $\psi$.  Note that actual pulsations tend to have
waveforms that are more complex than a sinusoidal function; however,
our general conclusions do not depend on such details, and
calculations can be modified to allow for any specific waveform
desired.

\begin{table}
\begin{center}
\caption{Simulation parameters for the stellar system.
\label{t:param}}
\begin{tabular}{ccccc}
\hline
$T_{\star,0}$  &  $\Delta$T & $r_{\star,0}$ & $\Delta r$ & $r_{\rm gas}$ \\
(K)&($T_{\star,0}$)&($r_{\sun}$)&($r_{\star}$)&($r_{\sun}$)\\
\hline\hline
10,000 & 0.20 & 1.8 & 0.20 & 3.6 \\
20,000 & 0.15 & 4.0 & 0.15 & 8.0 \\
30,000 & 0.10 & 6.5 & 0.10 & 13.0 \\
40,000 & 0.10 & 11.0 & 0.10 & 22.0 \\
\hline
\end{tabular}
\end{center}
\end{table}

\begin{figure}[ht]
\includegraphics[width=7.6cm]{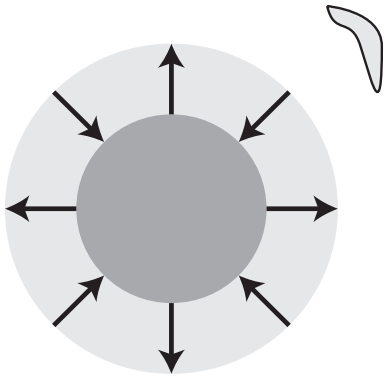}
\caption{Geometry of our model stellar system. The hypothetical
central star is a radially pulsating O or B~star (see text.)  As the
star expands and contracts, the surface temperature changes the
emitted blackbody spectrum, and thus varies the availability of UV
photons of different wavelengths. Hot, X-ray emitting gas with
temperature in the range of $10^6$ K and containing He-like ions is
located on-order 1~$r_\star$ above the surface.}
\label{f:geometry}
\end{figure}

In our model, the hot X-ray emitting gas is placed approximately
1~$r_\star$ above the surface of the star ({\em i.e.}, $r_{\rm gas}=2
r_\star$; see Fig.~\ref{f:geometry}). This hot zone is to be
considered as a ``test particle'' for our calculations.  A change in
temperature drives a change in the UV hardness of the radiation field
with time and thus the pumping of the $f$ level; a change in location
of the hot zone modifies the dilution factor $W(r)$ and consequently
the pumping owing to the mean intensity. If the hot zone moves
homologously with the surface of the star, such that the ratio
$r_\star/r_{\rm gas}$ remains constant, then $W$ does not vary. But if
the hot zone moves in some other fashion, or remains at a fixed
distance $r_{\rm gas}$, then $W$ will vary.  We model both moving
(oscillating) and fixed gas cases.

\begin{figure}[ht]
\includegraphics[width=7.6cm]{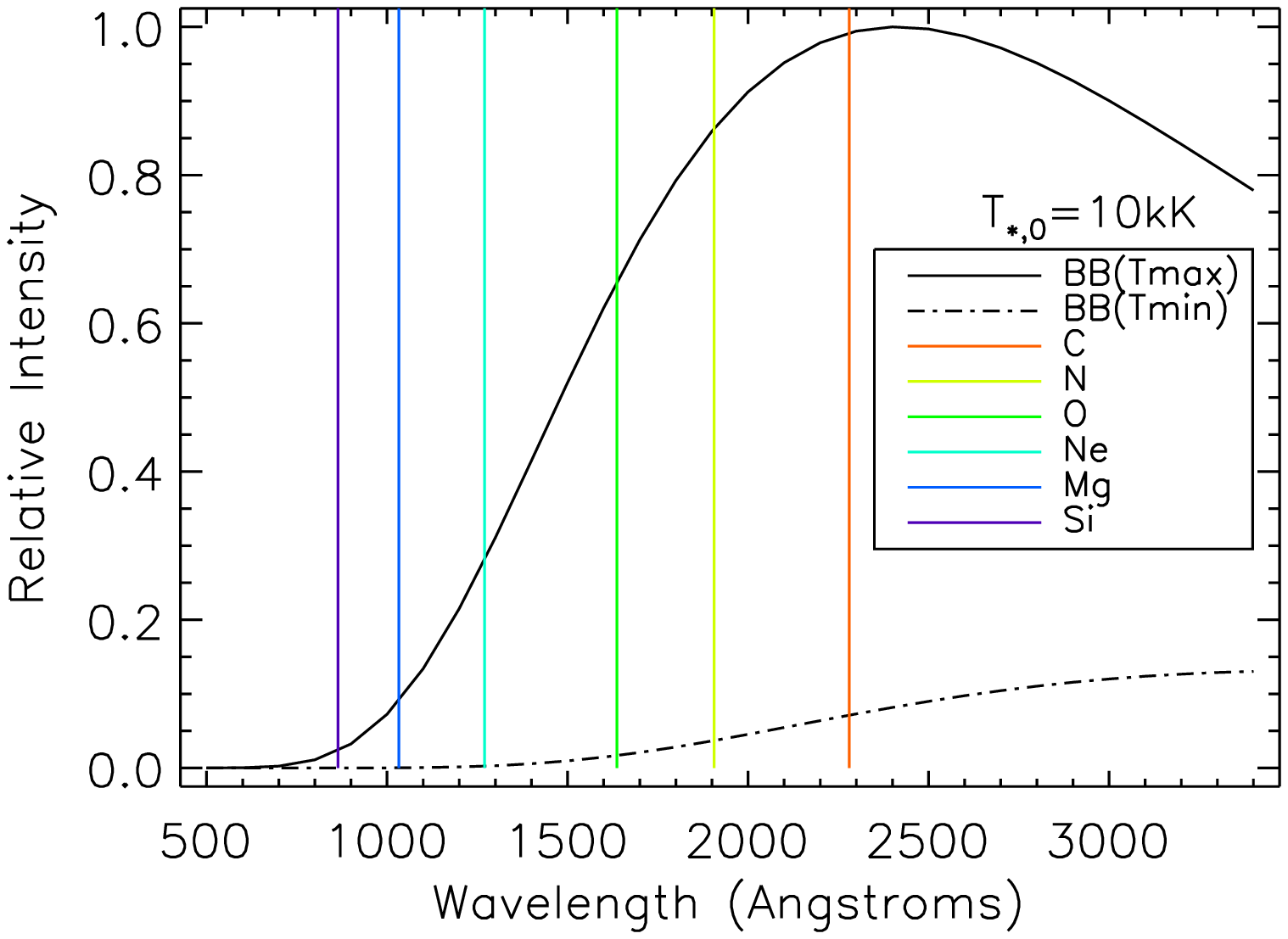}
\includegraphics[width=7.6cm]{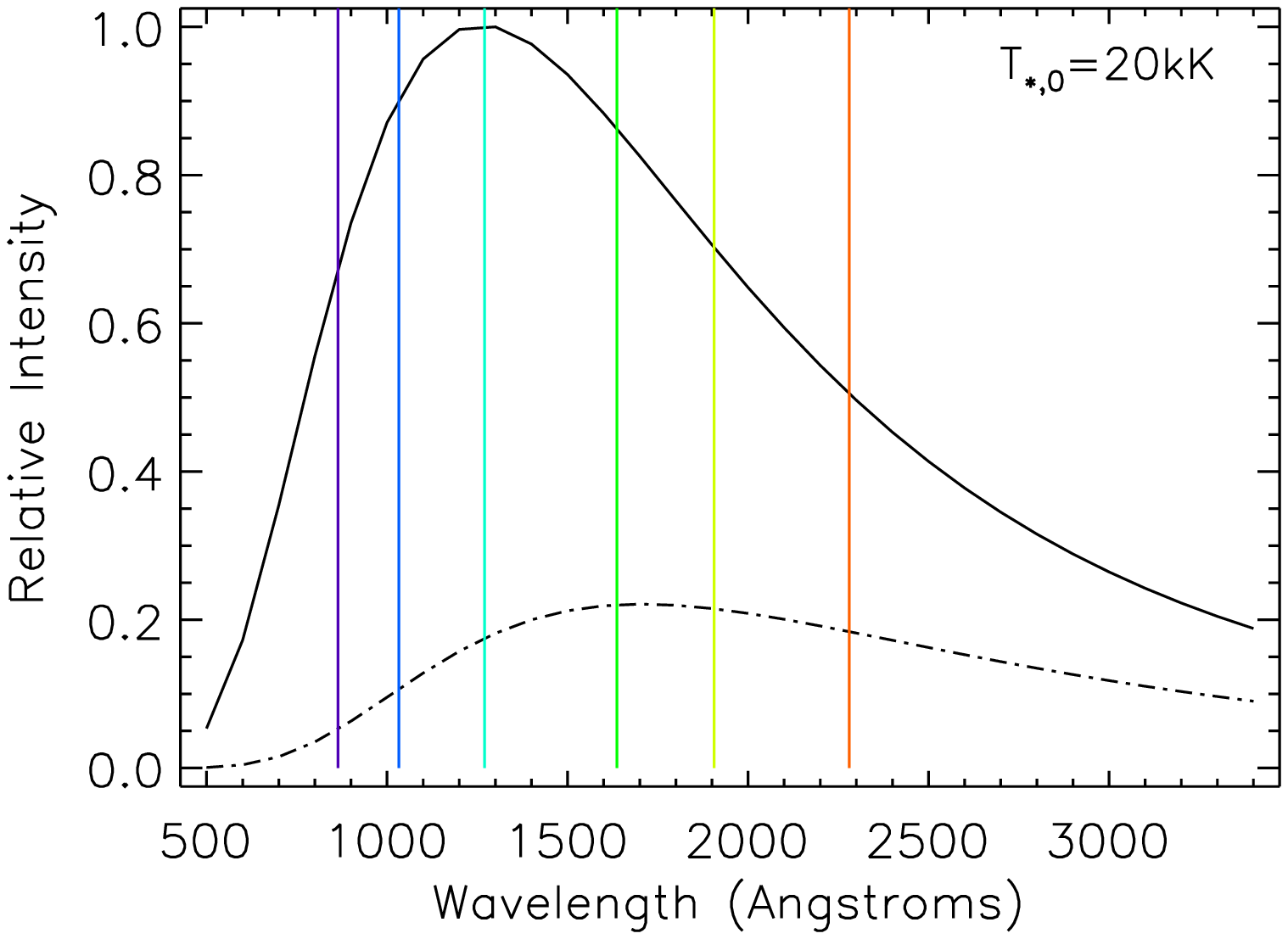} 
\includegraphics[width=7.6cm]{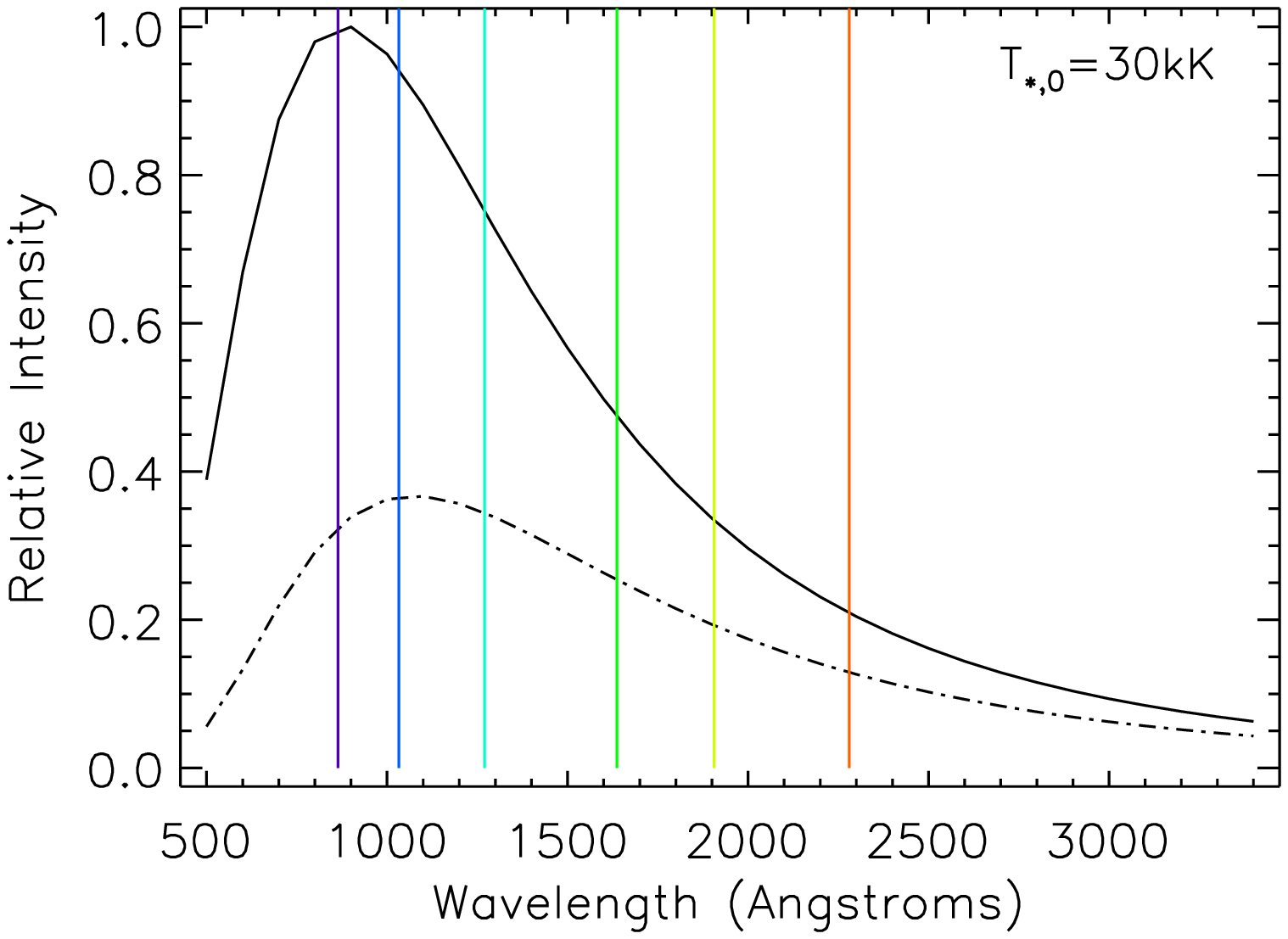} 
\includegraphics[width=7.6cm]{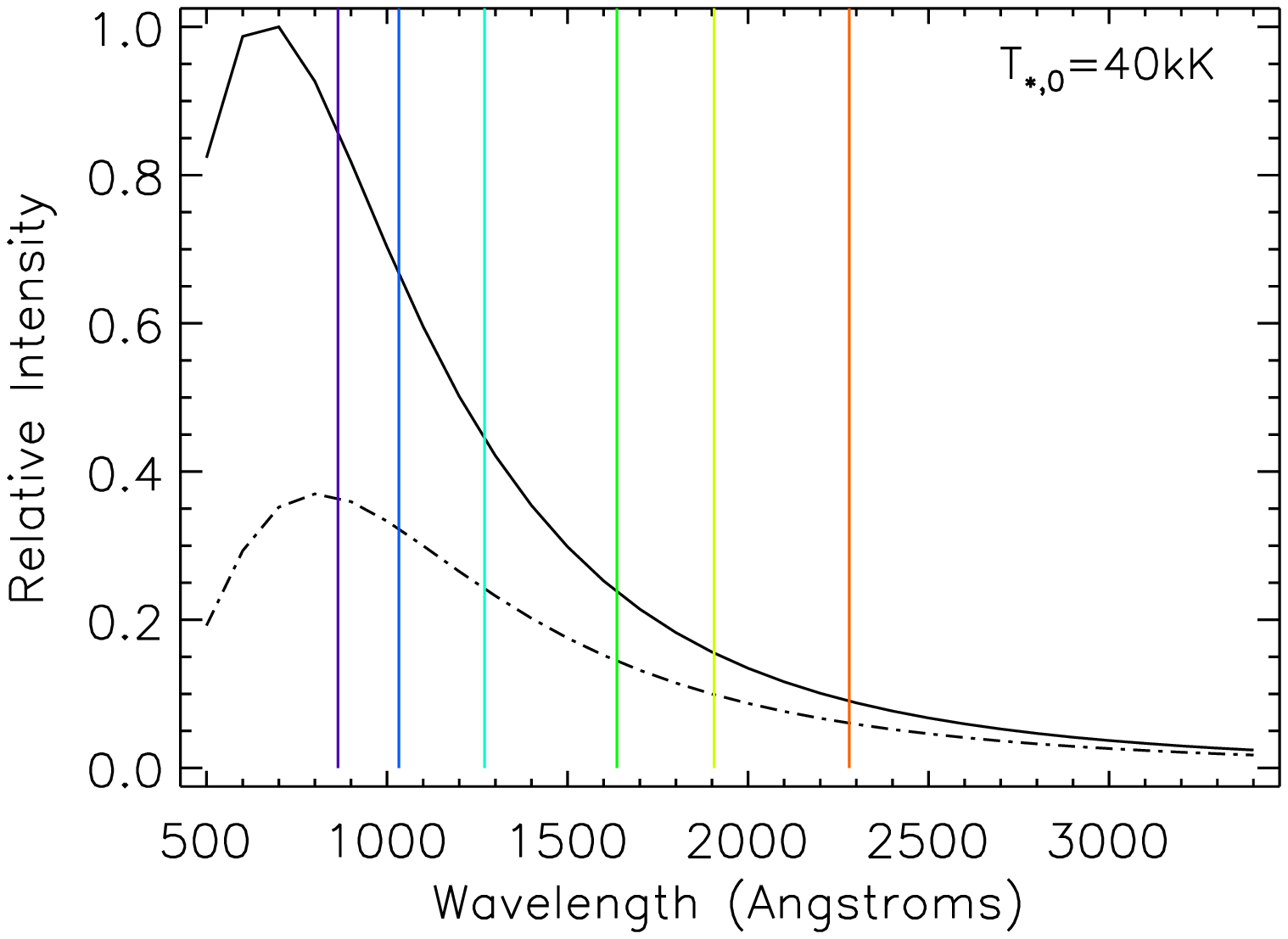} 
\caption{Blackbody spectra at maximum (solid line) and minimum (dot-dashed line) temperature over the
course of the pulsation for our hypothetical stars with average
effective temperature of 10,000, 20,000, 30,000 and 40,000~K (top to bottom).
UV-pumping wavelengths for 6 helium-like ions are also shown for
reference, \ion{C}{v} (red),
\ion{N}{vi} (yellow), \ion{O}{vii} (green), \ion{Ne}{ix} (teal), \ion{Mg}{xi} (blue) and
\ion{Si}{xiii} (purple).}
\label{f:BB}
\end{figure}

Finally, we calculate $R(t)$ for each of six commonly observed
helium-like ions: \ion{C}{v}, \ion{N}{vi}, \ion{O}{vii}, \ion{Ne}{ix},
\ion{Mg}{xi} and \ion{Si}{xiii} (see Table \ref{t:lines}). The energy
spacing between the intercombination and forbidden lines is different
for these elements, so each line will in general respond differently
to the spectral variations of the star throughout its pulsational
phase. The location of the pumping wavelengths of each ion in relation
to the shape of the blackbody curve for stars of different
temperatures is shown in Fig.~\ref{f:BB}.  In these plots, the black
lines show the Planck curve for the maximum and minimum temperature
the star reaches during its pulsation cycle.  The vertical colored
lines are at the wavelength of the photon required to move an electron
between the $f$ and $i$ levels in each He-like ion listed above.  The
figures give a sense of how pulsations in different stars can lead to
a diversity of responses in $R$ values as a function of the mean
stellar temperature.

\section{Results}
\label{s:results}

We have calculated the time-variability of $R$ for six helium-like
ions around three hypothetical radially-pulsating B~stars and one
low-massloss O~star. The main parameters of these systems are listed
in Table \ref{t:param}.  For our main discussion, we will present
simulations for X-ray emitting gas located at $r_{\rm gas}=2 r_\star$
as a fiducial value.  This model, with the gas at a discrete location,
may be applicable in some stellar systems, especially if the X-ray
emitting gas were constrained by magnetic fields.  The exact location
of the gas has implications for our results.  We therefore have
performed simulations with the gas at different distances. In general,
placing the gas closer to the surface of the star increases the
intensity of the radiation field uniformly, and therefore increases
the pumping in each ion and reduces the ratio $R$; increasing the
distance has the reverse effect.  In addition, because the radiation
field has less impact on $R$ in general when the gas is located
further away, any change in the radiation field will also have a
smaller effect on $R$ in an absolute sense.

\begin{figure}[htb]
\includegraphics[width=9cm]{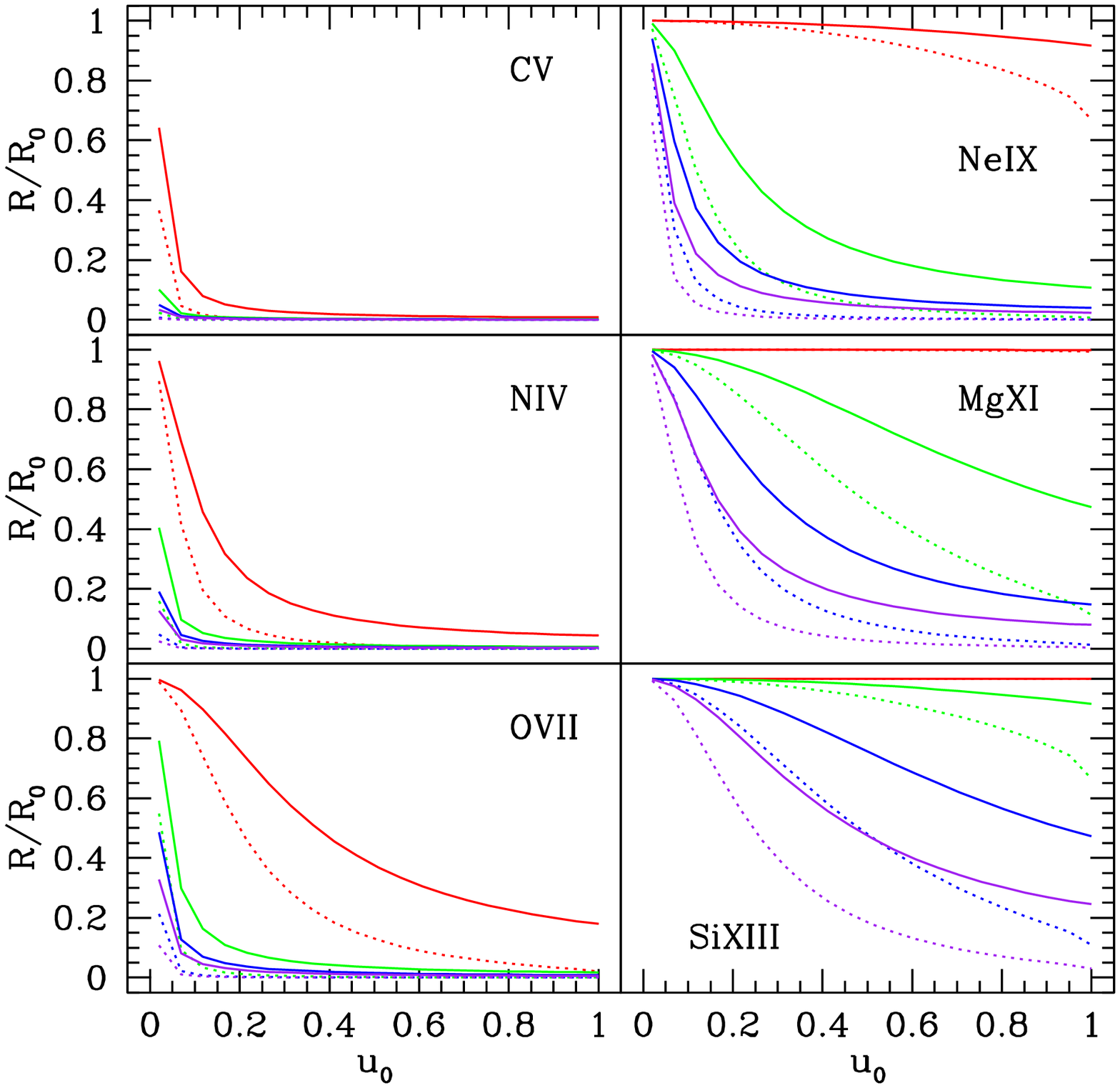}
\caption{Effects of the spatial distribution of X-ray emitting gas
on the behavior of $R$ for six helium-like ions: \ion{C}{v} (top left),
\ion{N}{vi} (middle left), \ion{O}{vii} (bottom left), \ion{Ne}{ix} (top right), \ion{Mg}{xi} (middle right) and
\ion{Si}{xiii} (bottom right) for each of our hypothetical stars with average
effective temperature of 10,000, 20,000, 30,000 and 40,000~K (red,
green, blue, and purple respectively). $R/R_0$ is plotted versus inner
radius of gas in inverse radius $u_0$. The substitution of variables,
$u=r_{\star}/r$, simplifies the calculation. The solid line indicates
how much $R$ is depressed from its nominal value if the gas is
distributed from $r=\infty$ ($u=0$) to $r=r_0$ ($u=u_0$).  The dotted
line shows the results if hot gas exists only at $u_0$. The value or
the dotted line at $u_0=1/2$ is equivalent to our fiducial value
elsewhere of $r_{\rm gas}=2 r_\star$ at zero displacement phase.}
\label{f:spatial}
\end{figure}

In the more general case, the hot gas may be distributed in radius
around the star.  Unfortunately, the exact distribution of gas is
highly model dependent, and the results will be strongly dominated by
the location of the inner radius.  (See \citet{leutenegger2006} for a
more in-depth discussion of this issue.)  We therefore calculated the
effect of all possible distributions on our
model. Fig.~\ref{f:spatial} plots $R/R_0$ versus location of the
gas. Here we use inverse distance from the star, $u=r_{\star}/r$, with
the inner radius of the X-ray emitting gas located at $u=u_0$. The
solid line shows the impact on $R$ if the gas is distributed from
$r_0$ to infinity ($u_0$ to 0), whereas the dotted line shows $R$ if
the location of gas is confined only to $u_0$. Note that the value of
the dotted line at $u_0=1/2$ is equivalent to $r_{\rm gas}=2 r_\star$.

This figure shows that there can indeed be a significant difference in
the $R$ ratio due to location and distribution of the gas. Of course,
the difference is generally larger the closer the inner radius of the
gas is to the surface of the star. The smaller radius increases both
the impact of a varying UV field on the closer material, and the total
amount of emitting material. In most cases, however, the difference
between gas confined at a single location and gas extending to
infinity is at most on order of the difference between two stellar
temperatures, \ie between $T_{\star,0}=20,000$K and
$T_{\star,0}=30,000$K.

\begin{figure*}[htb]
\includegraphics*[width=7.6cm]{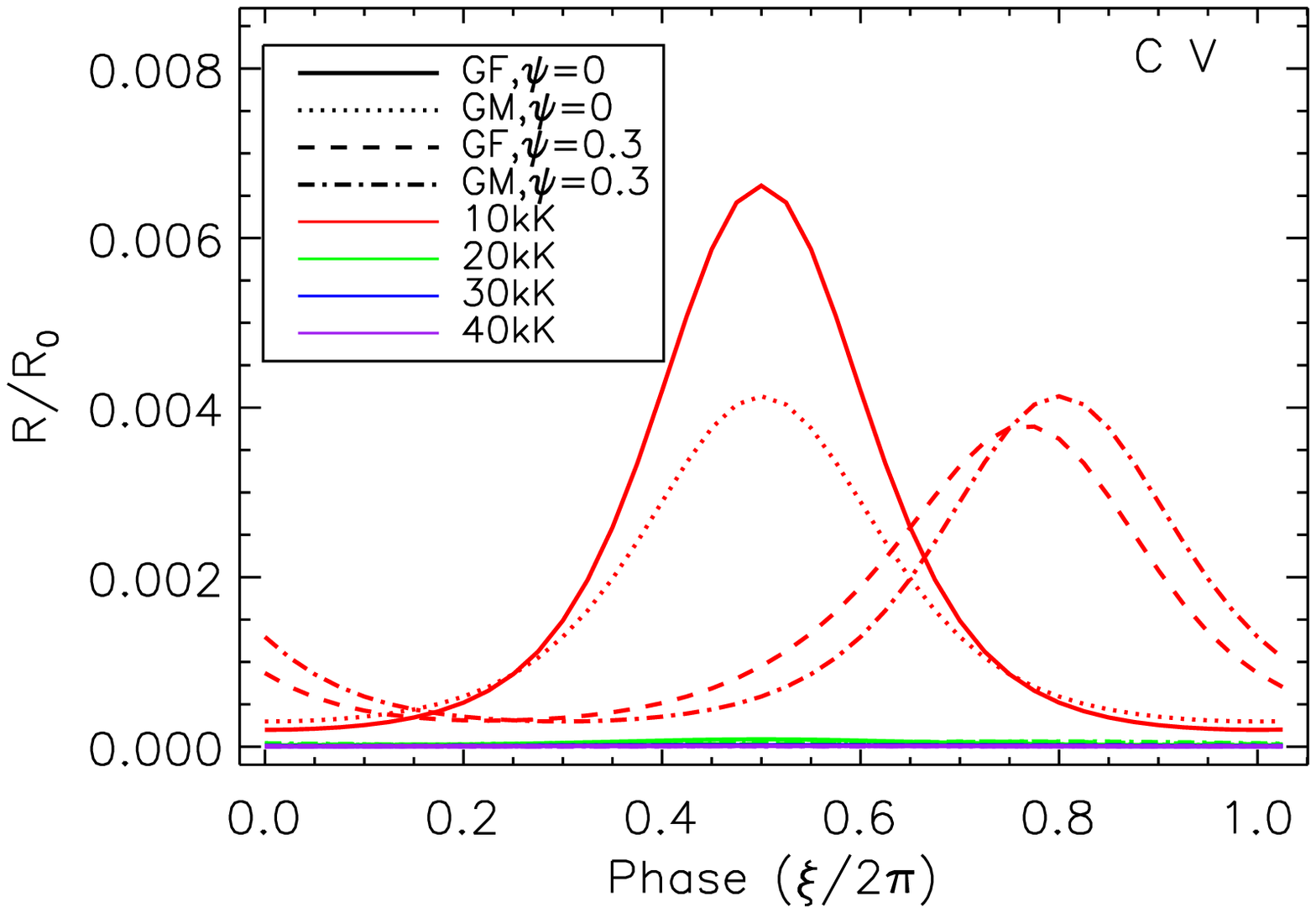}
\includegraphics*[width=7.6cm]{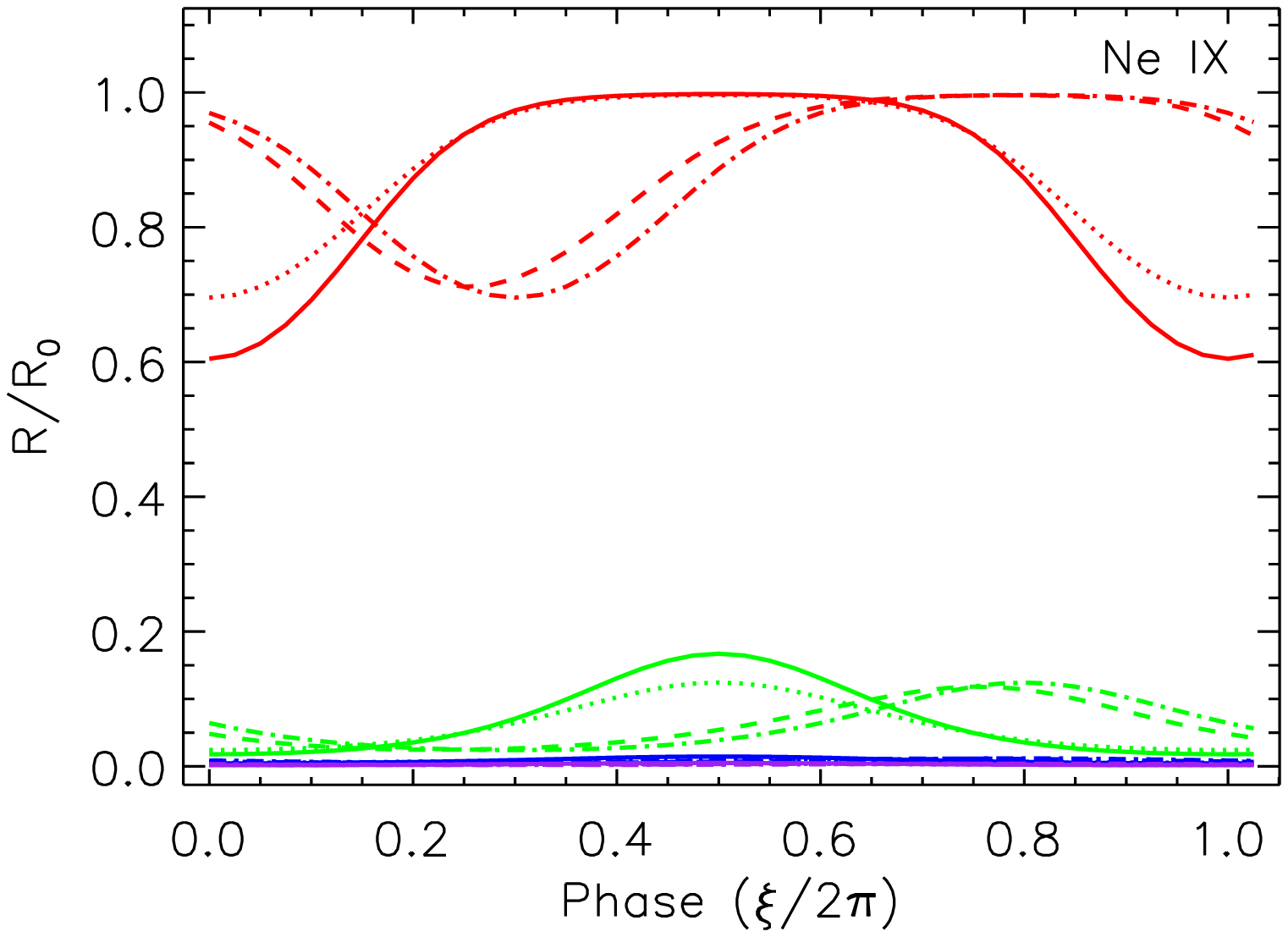}\\
\includegraphics*[width=7.6cm]{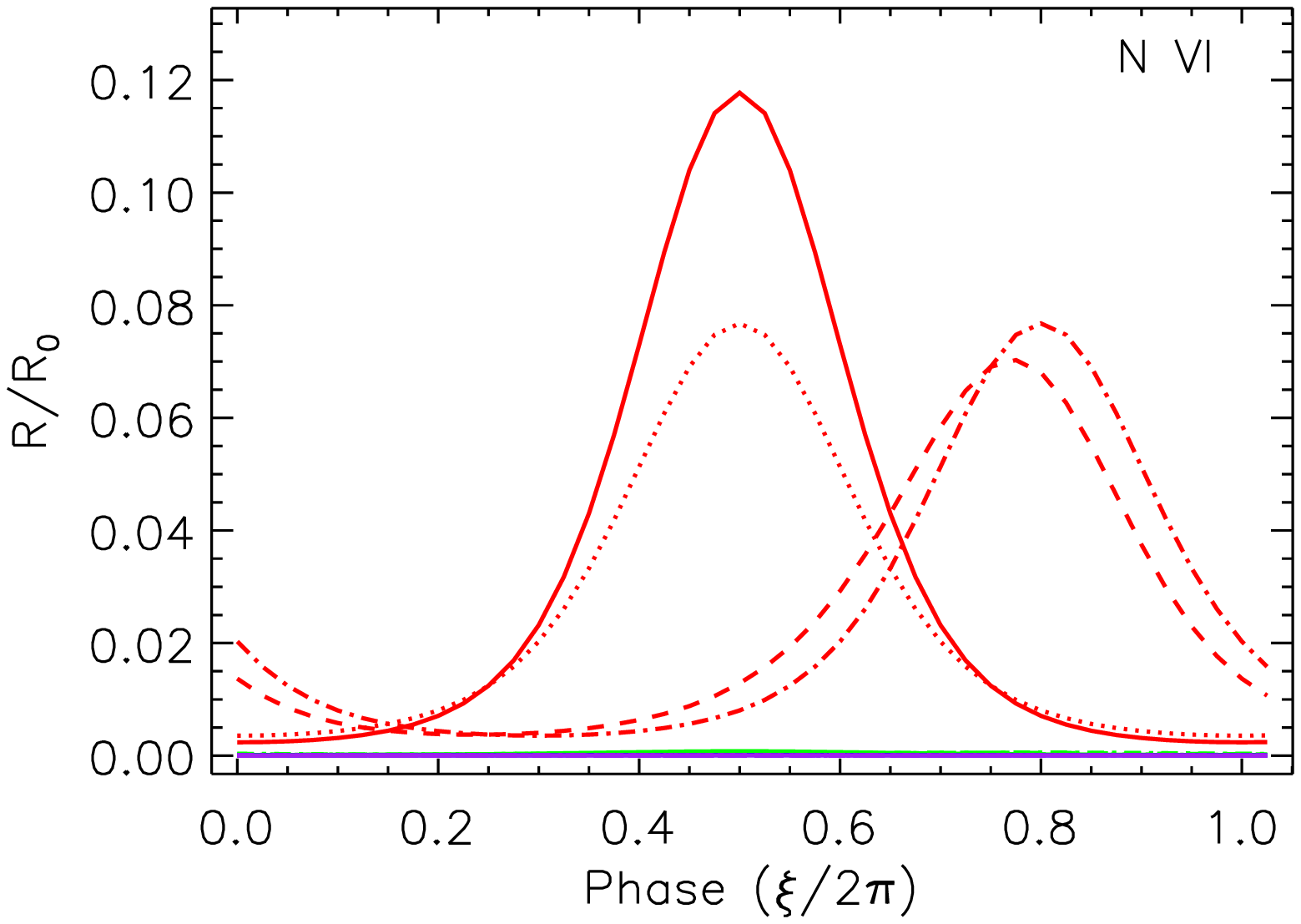}
\includegraphics*[width=7.6cm]{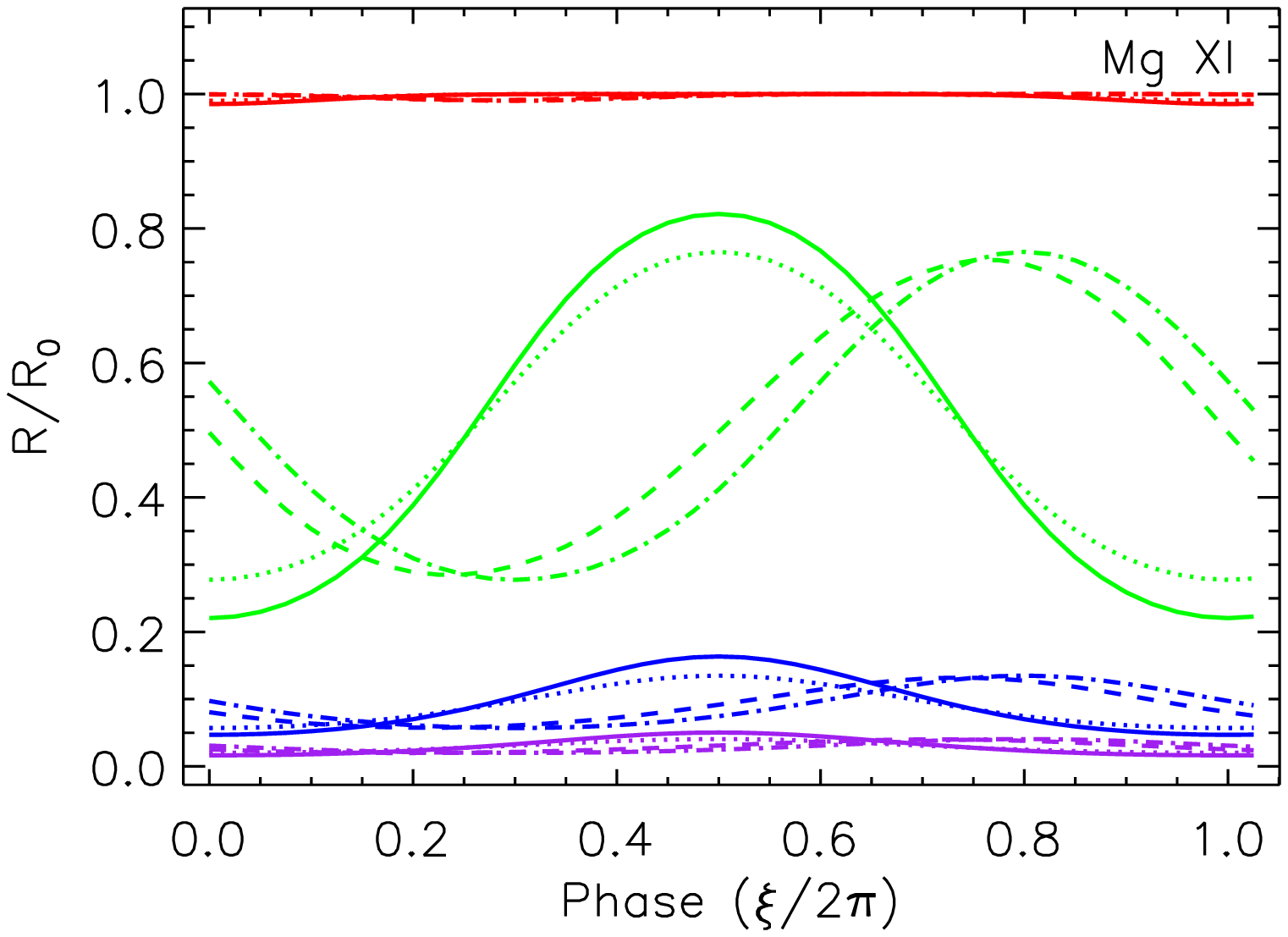}\\
\includegraphics*[width=7.6cm]{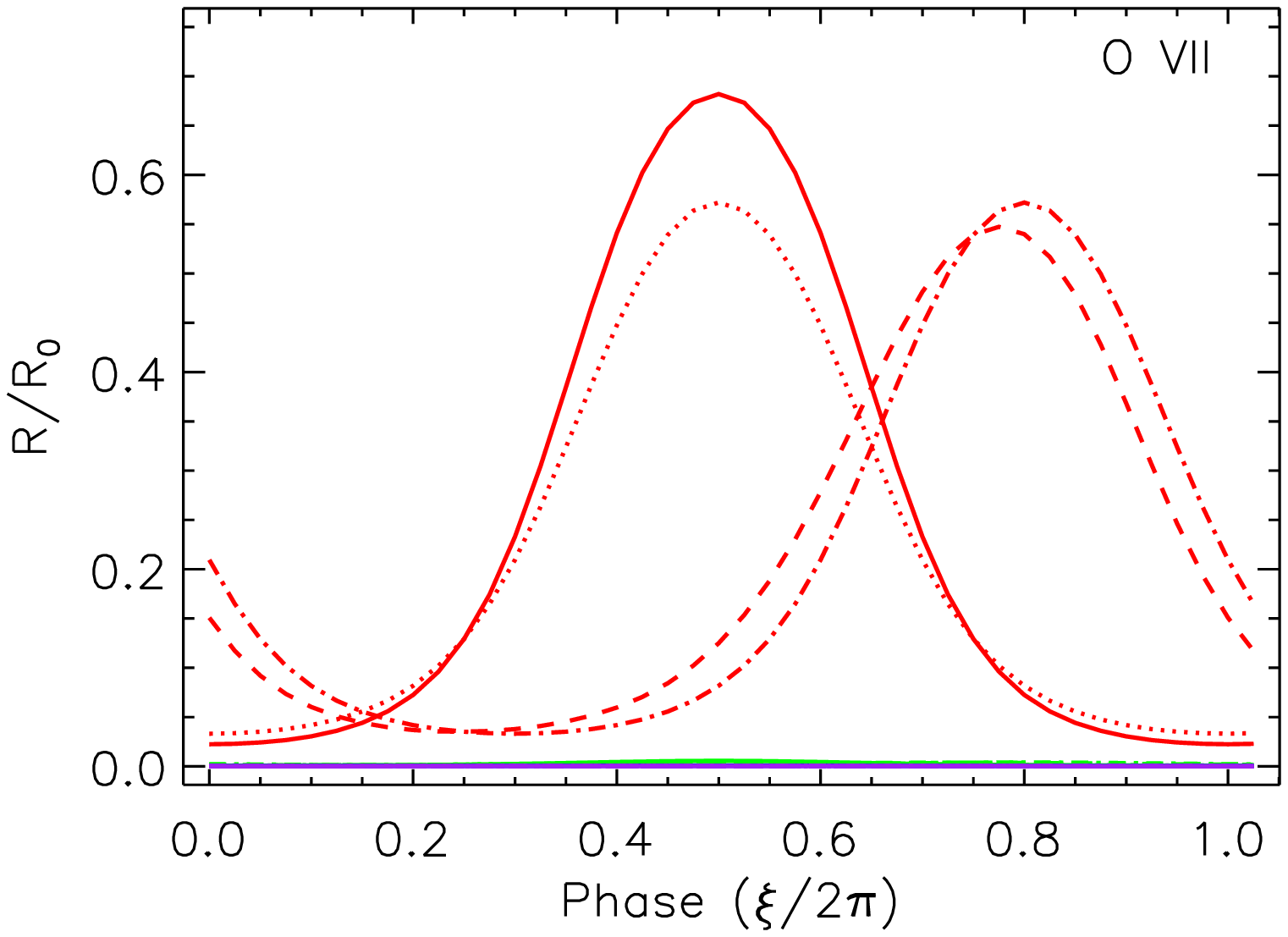}
\includegraphics*[width=7.6cm]{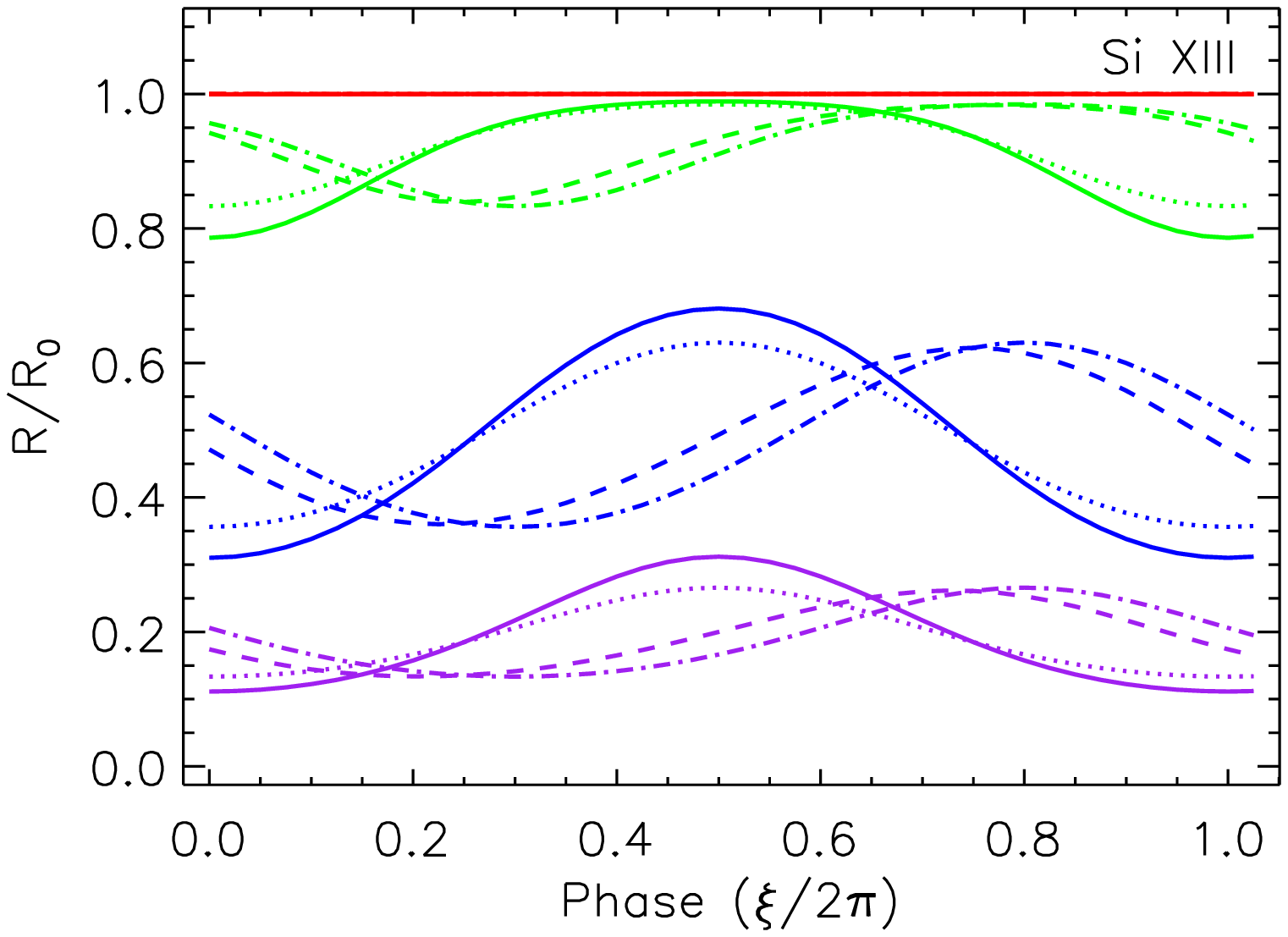}\\
\caption{Effects of temperature, phase offset, and gas motion on
the behavior of $R$ for $r_{\rm gas} = 2 r_\star$ for six helium-like
ions. The ions are \ion{C}{v} (top left), \ion{N}{vi} (middle left), 
\ion{O}{vii} (bottom left), \ion{Ne}{ix} (top right),
\ion{Mg}{xi} (middle right) and \ion{Si}{xiii} (bottom right).  
Simulation temperatures are 10,000, 20,000, 30,000 and 40,000~K (red,
green, blue, and purple respectively).  
``GF"  refers to simulations where
the gas remains at a fixed distance from the center of the star, while
``GM" signifies simulations where the gas moves in time and proportion
to the radial motion of the star. 
Thus the solid line is at fixed distance with no pulsation radial-temperature phase offset; the dotted line is moving gas with no offset; the dashed line is fixed gas with an offset; and the dot-dashed line is moving gas with a phase offset.
Note that the scale of the plots for
the lighter elements is reduced to show the range of variation, which
can be quite small.}
\label{f:comp}
\end{figure*}

Given the uncertainties inherent in modeling a spatial distribution,
from here on we return to the fiducial location of $r_{\rm gas}=2
r_\star$.  Our simulations investigate the effects of three factors:
temperature of the star, a phase offset between the radial and
temperature variations, and the cyclical variation in location of the
gas above the surface of the star.  We refer to these as the
temperature effect, phase effect, and W effect, respectively.
Fig.~\ref{f:comp} shows the variation of $R$ over time for each
combination of the these factors, for the six helium-like ions as
labeled.  Every ion does respond to some degree at each stellar
temperature. But the size of the response in each ion is quite
different. At a given temperature, the radiation field will either
almost completely suppress $R$ (in ions with lower energy UV-pumping
photons; note the scale of variation in Fig.~\ref{f:comp}), or have
little impact at all (in ions with greater separation between the
triplet energy levels.)  In the former case, $R$ itself will be hard
to measure observationally; in the latter, any variations will be
small and likewise difficult to detect.

\begin{figure}[ht]
\includegraphics[width=7.6cm]{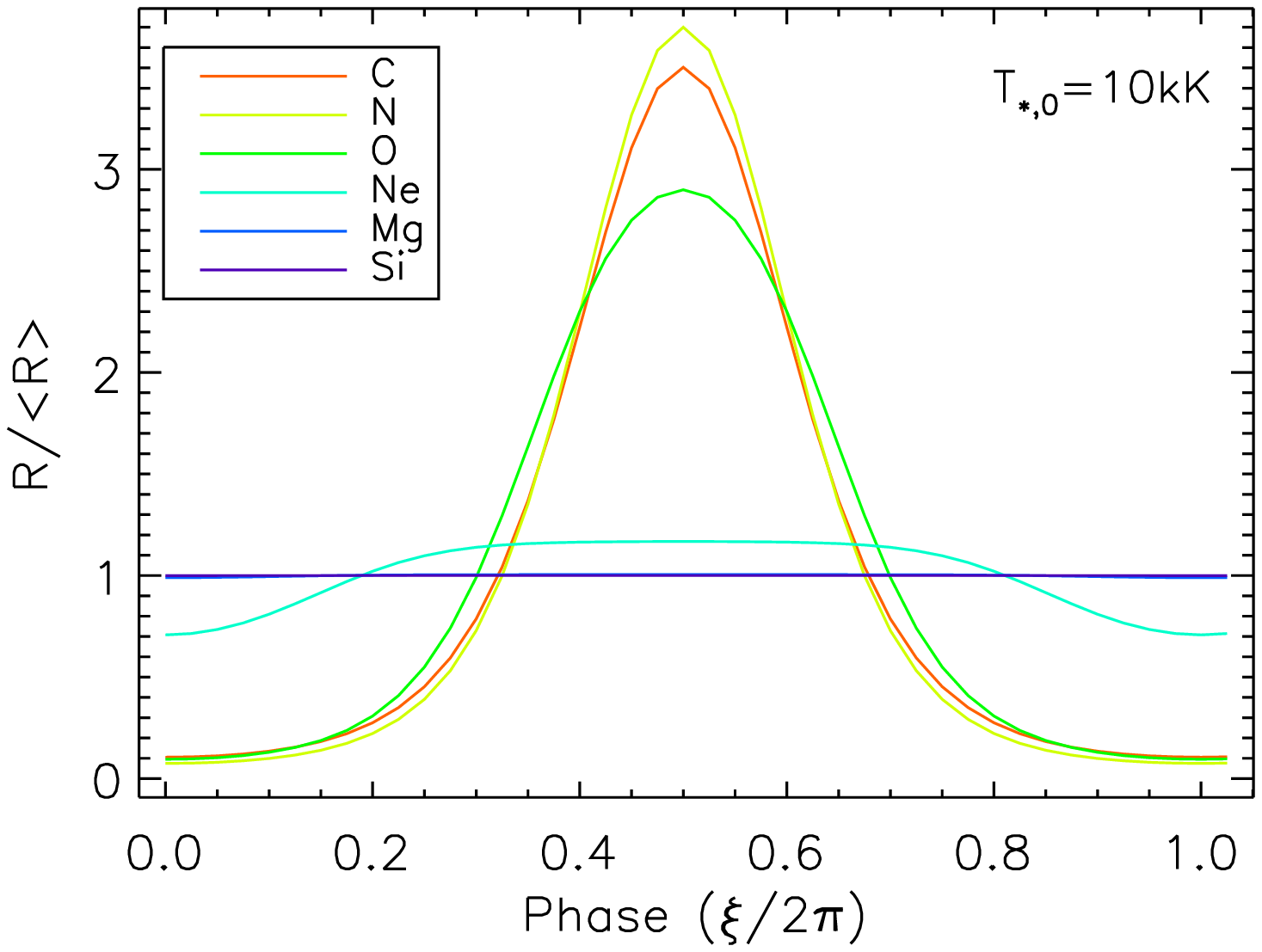}
\includegraphics[width=7.6cm]{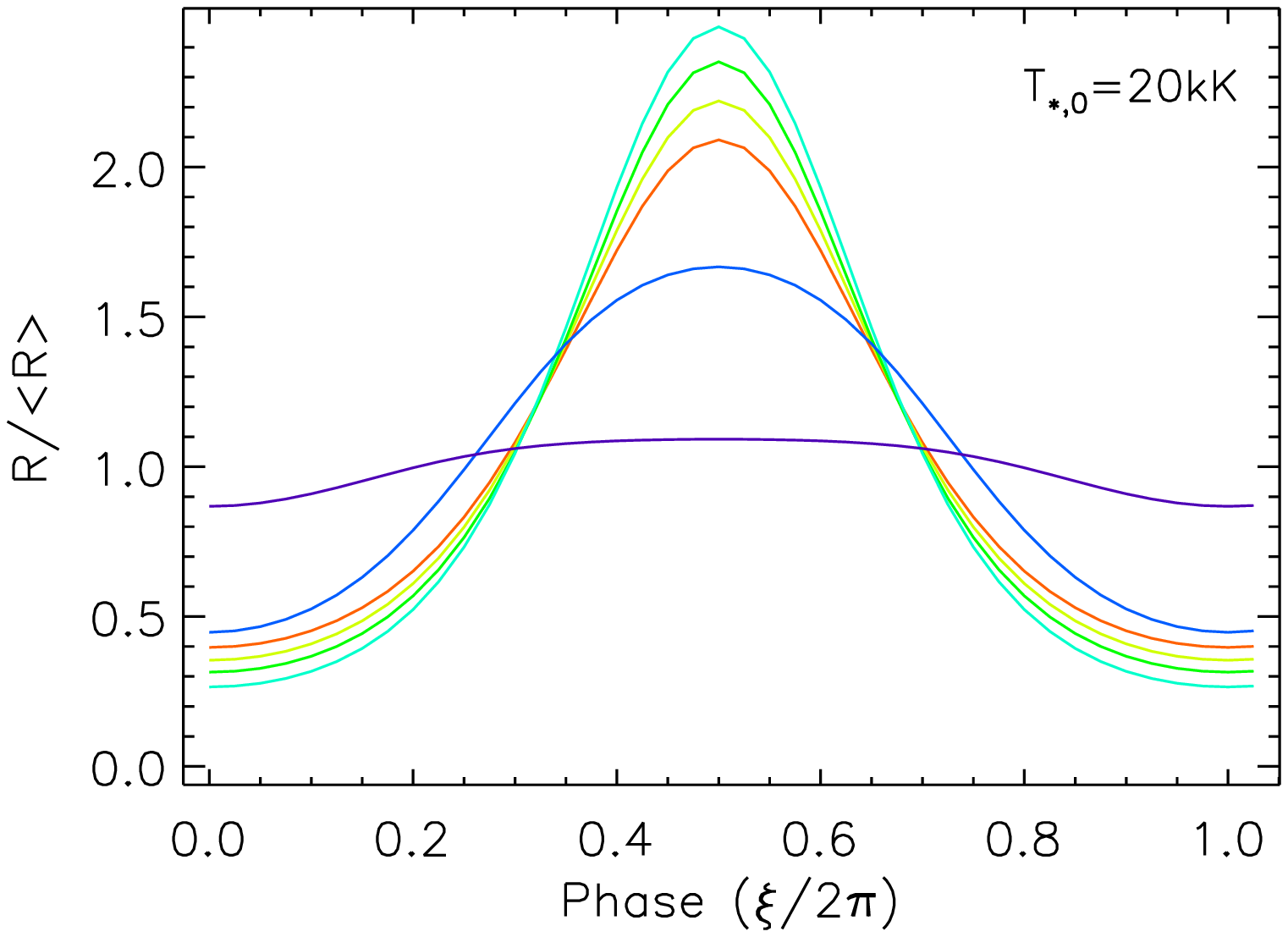} 
\includegraphics[width=7.6cm]{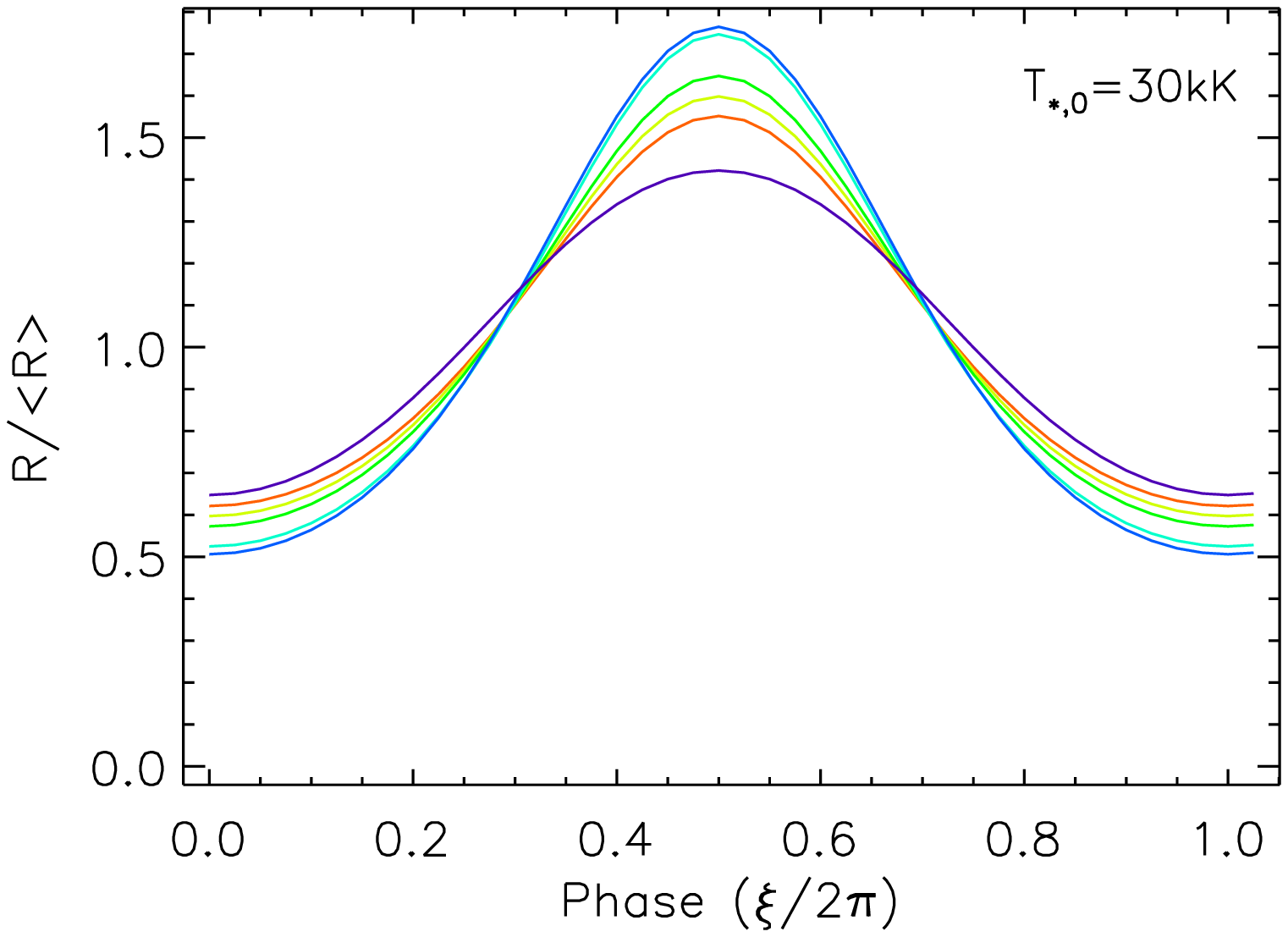} 
\includegraphics[width=7.6cm]{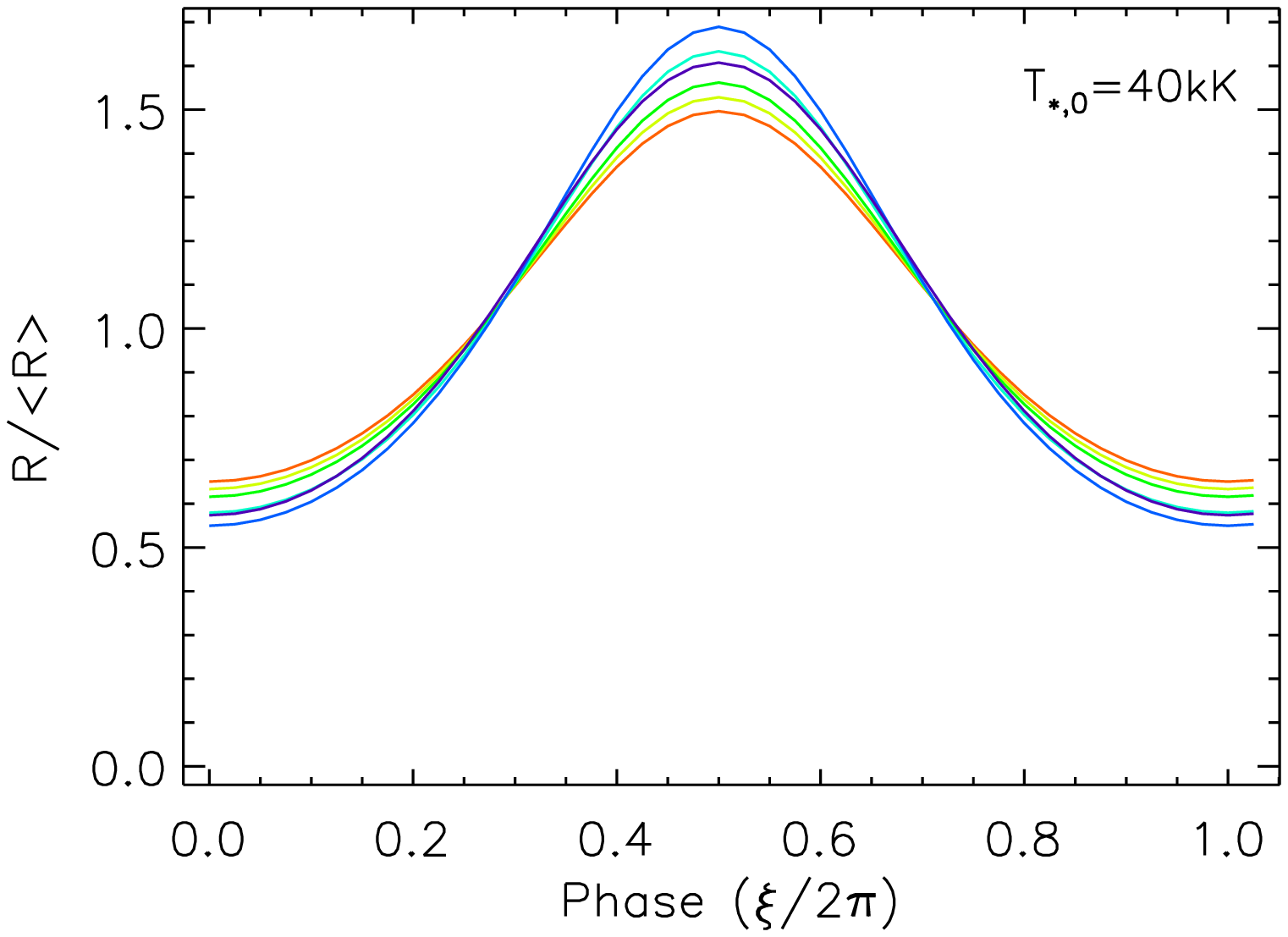} 
\caption{Variation in ratio $R/\left<R\right>$ for each of the six
helium-like ions, \ion{C}{v} (red);
\ion{N}{vi} (yellow); \ion{O}{vii} (green); \ion{Ne}{ix} (teal); \ion{Mg}{xi} (blue); and
\ion{Si}{xiii} (purple), 
for our model for a star with average
effective temperatures 10,000, 20,000, 30,000 and 40,000~K (top to
bottom).  Note that the absolute scale of variation changes
dramatically -- see the values of $R_0$ and $\left<R\right>$ for each
ion in Table \ref{t:stats}. This figure shows the differences in the
width and height of the peaks and troughs in the periodic variation of
the ratio.}
\label{f:R_Rave}
\end{figure}

At each temperature there are one or two ions where the pumping
frequency is in the right portion of the stellar spectrum to
substantially affect the level populations but not totally depopulate
them. In these ions, changes in the radiation field may have a
significant, potentially observable effect on $R$.
Table~\ref{t:stats} shows the statistics of variation for each ion for
each of our stellar temperature models in the fixed-gas,
no-phase-offset simulations.  The first column shows $R_0$ from
\citet{blumenthal1972}, the fiducial value that all changes are
relative to. $\left<R\right>$ is the time-average value of $R$ found
in each simulation.  $\sigma_R$ is the standard deviation of the
variation in $R$ with phase. If the shape of the $R$ curve was
sinusoidal, in the form $R=a+b \sin(\xi)$, then $a=\left<R\right>$ and
$b=\sigma_R$. But as Fig.~\ref{f:R_Rave} shows, $R$ is not always
sinusoidal, and in those cases these values have a different
relationship to the overall behavior with time.  The fifth column of
Table~\ref{t:stats} shows how much the value of $R$ has been depressed
from the nominal value (the case of no radiative pumping), while the
sixth column shows the scale of variation relative to the average
value. These two columns provide a sense of the observability of the
ratio and the potential to detect the variations in the ratio,
respectively.

These results are presented in another way in Fig.~\ref{f:comp}.
These plots also demonstrate the impact of a phase offset between the
temperature and radius variations in the star, and of oscillating
versus fixed gas. If $\psi=0$, the maximum in $R$ always occurs at the
same phase as the minimum in star temperature, whether the gas is
moving in time with the surface of the star or not.  This is because
the number of available UV photons is mostly set by the temperature of
the star, and if the gas is moving with the star, the change in radius
has no effect on the radiation field at the gas location. If, however,
the gas is fixed and the star moves outward at the same time the
temperature increases, both factors work to increase the intensity of
the radiation field.

In the non-synchronous variability case ($\psi\ne0$), if the emitting
gas is moving with the radius of the star, the line ratios will have
similar properties to the synchronous case, but with an offset in
phase. This is because changes in $R$ are almost completely due to
changes in the temperature of the star.  If the gas is fixed, on the
other hand, the effect is the reverse of the $\psi=0$ case --
diminishing the change in $R$ rather than increasing it, because the
increase in intensity from being closer to the surface of the star
works to counteract the increase in intensity due to a higher stellar
temperature. The changes in the temperature still dominate, however.
If $\psi=0.5$, the maximum $R$ will still occur at the same time as
the minimum $T_\star$; at any other $\psi\ne0$ offset, the maximum $R$
will occur just after the minimum stellar temperature, as the stellar
surface begins to approach the gas.

\begin{table*}
\begin{center}
\caption{Model results for for the three stars listed in Table
\ref{t:param} with no temperature-radius phase offset and fixed gas
location.
\label{t:stats}}
\begin{tabular}{cccccc}
\hline
Ion & $R_0$ $^1$
& $\left<R\right>$ & $\sigma_R$ & 
$\left<R\right>/R_0$& $\left<R\right>/\sigma_R$ \\
\hline\hline
\multicolumn{6}{c}{$T_{\star,0}=10,000$K}\\
\hline
\ion{C}{v} & 11.3\poh & 0.02 & \poh0.02 & $<$0.01 & 0.89 \\
\ion{N}{vi} & \poh5.13 & 0.16 & \poh0.20 & \poh0.03 & 0.83 \\
\ion{O}{vii} & \poh3.85 & 0.91 & \poh0.92 & \poh0.24 & 0.98 \\
\ion{Ne}{ix} & \poh3.17 & 2.71 & \poh0.48 & \poh0.85 & 5.65 \\
\ion{Mg}{xi} & \poh3.03 & 3.02 & \poh0.02 & \poh1.00 & $>$100 \\
\ion{Si}{xiii} & \poh2.51 & 2.51 & $<$0.01 & \poh1.00 & $>$100 \\
\hline
\multicolumn{6}{c}{$T_{\star,0}=20,000$K}\\
\hline
\ion{C}{v} & 11.3\poh & $<$0.01 & $<$0.01  & $<$0.01  & \poh1.68 \\
\ion{N}{vi} &  \poh5.13 & $<$0.01  & $<$0.01  & $<$0.01  & \poh1.53 \\
\ion{O}{vii} & \poh3.85 & $<$0.01 & $<$0.01  & $<$0.01 & \poh1.41 \\
\ion{Ne}{ix} & \poh3.17 & \poh0.21 & \poh0.17 & \poh0.07 & \poh1.30 \\
\ion{Mg}{xi} & \poh3.03 & \poh1.49 & \poh0.68 & \poh0.49 & \poh2.18 \\
\ion{Si}{xiii} & \poh2.51 & \poh2.27 & \poh0.19 & \poh0.91 & 11.8\poh \\
\hline
\multicolumn{6}{c}{$T_{\star,0}=30,000$K}\\
\hline
\ion{C}{v} & 11.3\poh &$<$0.01  &$<$0.01 & $<$0.01  & 3.00\\
\ion{N}{vi} &  \poh5.13 &$<$0.01 &$<$0.01 &$<$0.01  & 2.79 \\
\ion{O}{vii} & \poh3.85 & $<$0.01  &$<$0.01  & $<$0.01 & 2.60 \\
\ion{Ne}{ix} & \poh3.17 & \poh0.03 & \poh0.01 & $<$0.01 & 2.29 \\
\ion{Mg}{xi} & \poh3.03 & \poh0.28 & \poh0.13 & \poh0.09 & 2.22 \\
\ion{Si}{xiii} & \poh2.51 &\poh1.20 & \poh0.34 & \poh0.48 & 3.51 \\
\hline
\multicolumn{6}{c}{$T_{\star,0}=40,000$K}\\
\hline
\ion{C}{v} & 11.3\poh &$<$0.01  &$<$0.01 & $<$0.01  & 3.29\\
\ion{N}{vi} &  \poh5.13 &$<$0.01 &$<$0.01 &$<$0.01  & 3.11 \\
\ion{O}{vii} & \poh3.85 & $<$0.01  &$<$0.01  & $<$0.01 & 2.95 \\
\ion{Ne}{ix} & \poh3.17 & $<$0.01 & $<$0.01 & $<$0.01 & 2.65 \\
\ion{Mg}{xi} & \poh3.03 & \poh0.09 & \poh0.04 & \poh0.03 & 2.45 \\
\ion{Si}{xiii} & \poh2.51 &\poh0.49 & \poh0.18 & \poh0.19 & 2.68 \\
\hline
\end{tabular}
\tablebib{
(1)~\citet{blumenthal1972}.
}
\end{center}
\end{table*}

\begin{table*}
\begin{center}
\caption{
Some values of scales of $R/\delta R$ in a sample of stars with
observed $R$ ratios from the literature. This measure shows the
potential observability of variability of the ratio in different lines
for these stars, not a variation that has been detected.
\label{t:obs}}
\begin{tabular}{ccccccccc}
\hline
Object & \ion{S}{xv}&\ion{Si}{xiii}&\ion{Mg}{xi} &\ion{Ne}{ix}
&\ion{O}{vii}& \ion{N}{vi} & \ion{C}{v}& Reference\\
\hline\hline
$\zeta$~Pup& &4.50&4.00&2.92&&&&1 \\
$\zeta$~Ori& &1.19&2.25&1.50&&&&1 \\
 $\xi$~Per& &0.154&1.27&1.00&&&&1 \\
 $\zeta$~Oph& &1.125&3.00&1.00&&&&1 \\
 $\tau$~Sco& &3.00&3.46&&5.60&3.09&1.48&2\\
 $\tau$~Sco&1.88&7.78&6.86&$R<$0.03&$R<$0.13&&&3\\
\hline
\end{tabular}
\end{center}
\tablebib{
(1)~\citet{oskinova2006}; (2) \citet{mewe2003a}; (3) \citet{cohen2003}.
}
\end{table*}

\begin{figure}[ht]
\includegraphics[width=7.6cm]{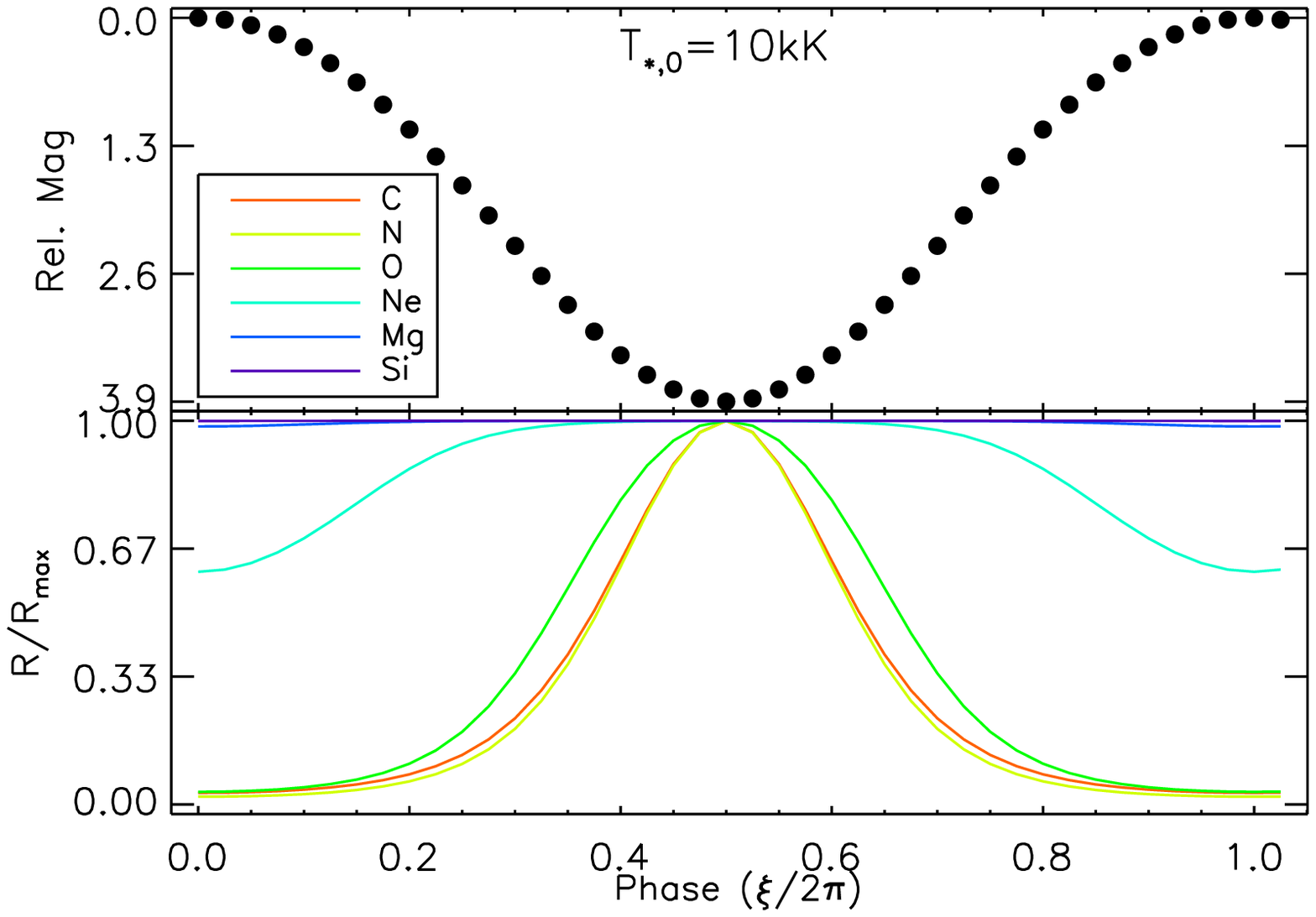}
\includegraphics[width=7.6cm]{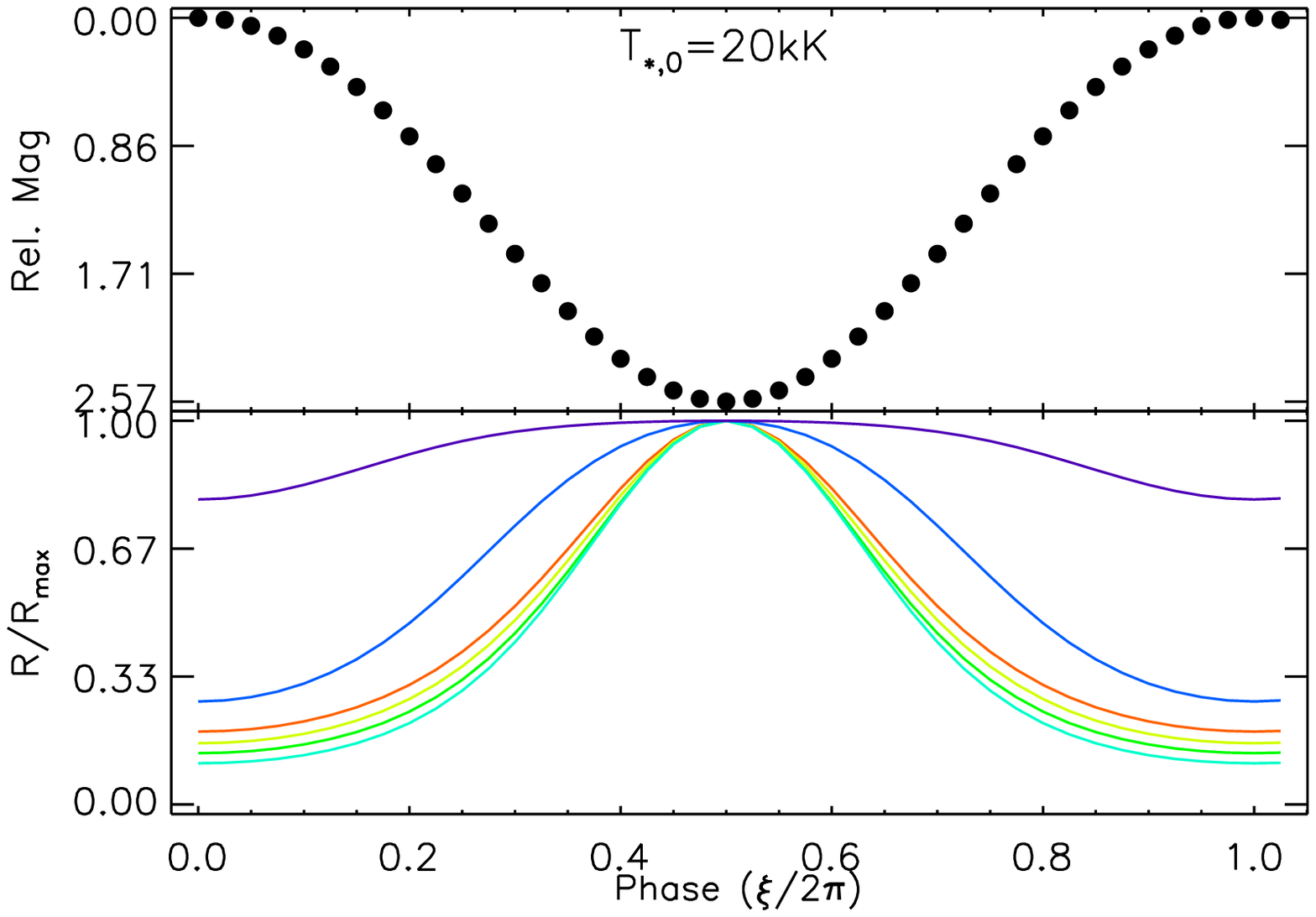} 
\includegraphics[width=7.6cm]{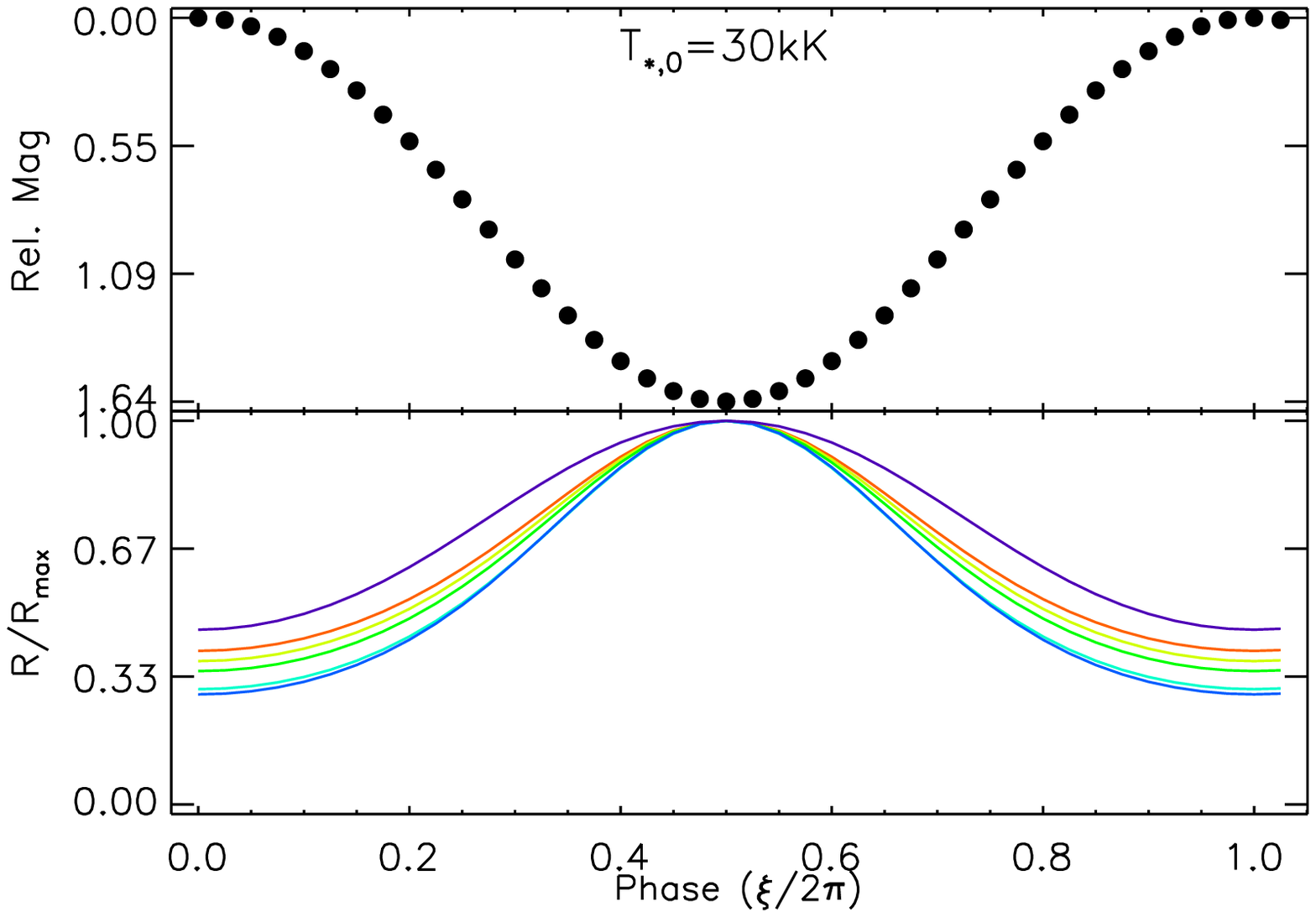} 
\includegraphics[width=7.6cm]{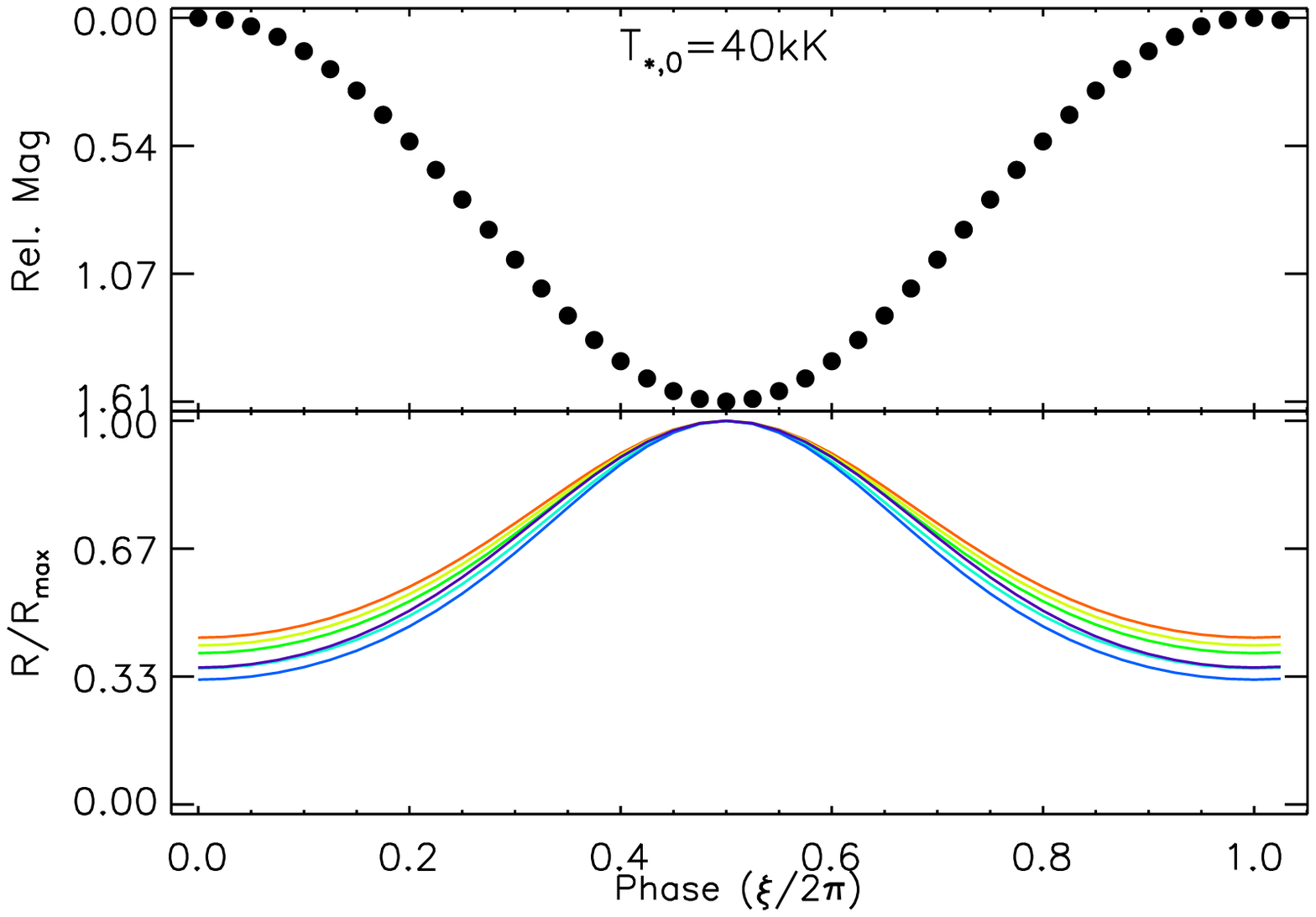} 
\caption{Variation in $R$ versus approximate visible magnitude light
curve (dots; see text) for six helium-like ions: \ion{C}{v} (red),
\ion{N}{vi} (yellow), \ion{O}{vii} (green), \ion{Ne}{ix} (teal), \ion{Mg}{xi} (blue), and
\ion{Si}{xiii} (purple); for stars with
average effective temperatures 10,000, 20,000, 30,000 and 40,000~K
(top to bottom). Note that the peak in $R$ corresponds with the
minimum optical brightness.}
\label{f:R_vary}
\end{figure}

Fig.~\ref{f:R_Rave} shows the difference in response of $R$ to stellar
pulsations for the different ions in the fixed-gas, no-phase-offset
case. Note that the absolute scale of variation changes dramatically
between ions -- see Table~\ref{t:stats} for values of $
\left<R\right>$ and $\sigma_R$. In heavy ions at lower temperatures,
there are few photons at the wavelengths needed to depopulate the
forbidden line level, and consequently $R\sim R_0$. As the temperature
increases, more depopulating photons become available, and $R$ varies
with phase. Eventually even the heaviest ions have enough photons to
saturate the response, and the profiles of all ions show the same
shape, though the absolute scales still differ. This divergence from
sinusoidal response at moderate intensity impacts the root mean square
of the value of $R$ of each ion, and thus the expected value for $R$
if observed in this regime.  Fig.~\ref{f:R_vary} demonstrates these
trends and shows the behavior of $R$ for each of our six ions in
relation to the approximate visible light curve of the pulsating star.
The approximate visible magnitude here is calculated by finding the
total flux from the blackbody spectrum in a 64 nm band centered on 540
nm.

Table~\ref{t:obs} gives a sense of the current state of observational
possibilities for the $R$ ratio in actual hot stars.  A comparison
between Table~\ref{t:obs} and Table~\ref{t:stats} gives a sense of how
observable the variations in our simulations may be with current
instruments. In the observational table, $R/\delta R$ represents the
data quality typical of current observations.  It is the value of the
measured ratio $R$ divided by the uncertainty in that measurement;
thus a value of 1 indicates uncertainty equivalent to the measurement,
while a larger number means greater certainty and a greater chance of
detecting variability.  The value is listed for different triplet
lines for different stars, as indicated.  The star $\tau$~Sco appears
twice: once for {\em XMM-Newton} spectra (first) and once for {\em
Chandra} data (second). Note that most of the sources listed in
Table~\ref{t:obs} are early-type O~stars, whereas our models are for B
and late-type O~stars.  However, the data quality are shown primarily
as a guide for the scope of what amplitude levels may be reasonably
{\em detected} as compared to predicted amplitudes of variations.

$\left<R\right>/\sigma_R$ is the equivalent value from our simulations
(Table~\ref{t:stats}), yet here the implication for observability is
different. If $\left<R\right>$ is small, the total flux will be low
and therefore hard to measure, requiring a small signal-to-noise ratio
and long exposure times.  Further, if $\sigma_R$ is small, the change
in $R$ at any time will be small, and harder to detect.  Thus to be
detectable, we need an $\left<R\right>\gtrsim0.1$ and $\sigma_R \sim
\left<R\right>$.

To summarize the detection possibilities, the only ions in our
simulations that seem to be candidates for such detection are
\ion{Ne}{ix} for the $T_{\star}=10,000$K star and \ion{Mg}{xi} and
\ion{Si}{xiii} for the $T_{\star}=20,000$ and 30,000K stars. It might
be possible, though hard, to detect variation in \ion{Si}{xiii} around
a $T_{\star}=40,000$K star.  Though our simulations are coarsely
gridded, these results are likely to be robust, since the only element
between \ion{Ne}{ix} and \ion{Mg}{xi} is \ion{Na}{x}, which is not
commonly seen in observations. Bear in mind that these simulations
were designed for maximum variability and detectability.  A variety of
factors, discussed in the next section, could make actual detection
more difficult.

\section{Discussion}
\label{s:discussion}
%

We have investigated variability in $R$ for He-like ions in massive
stars, caused by cyclical changes in the radiation field of a star. We
have made several simplifications that allow us to explore the upper
limits of the possible effect of stellar pulsation on observed
variability in $R$: we have assumed a hot O or B~star with purely
radial pulsation to maximize the influence of radiative pumping
effects on X-ray diagnostic $R=f/i$.  We have chosen a B or low-mass
loss O~star, so that we could plausibly assume that the electron
density in the gas is much lower than the critical density, so that
$R$ is determined solely by the stellar radiation field. And we have
assumed an X-ray emitting gas with a range of temperatures $\sim10^6 -
10^7$ K, to isolate the effect of the radiation field from more
complicated effects of changes in $R_0$, $T_X$, and ionization
fraction within the gas.  We have also confined the emitting gas to a
single location, which isolates the response to the local UV radiation
field rather than integrating over many locations and UV flux
densities.  This model might be applicable for an early B or late
O~star with a strong magnetosphere to magnetically channel or confine
the hot gas near the stellar surface.

With these optimizing assumptions, our model predicts that a variation
in the $R$ ratio could be detectable in one or two lines
simultaneously.  This is because, with a reasonable blackbody spectrum
variations changing UV flux, the value of $\phi_c$ and the pumping
wavelength have sufficient dynamic range that most lines are in one or
the other limiting case: virtually no pumping or extreme pumping. Our
model predicts which lines should be variable for stars of different
surface temperatures.  Thus observations of this kind could be used to
probe the environment of pulsators or stars with magnetically confined
wind shocks. $\beta$ Cephei, a B-type star at a distance of 182 pc,
could be an interesting target for such analysis. The prototype for
$\beta$ Cepheid class variables, it has known variability, with a main
pulsation mode period of 274 minutes and other non-radial modes
\citep{favata2009}.

If variability were observed in $R$ in several lines at once, however,
our model makes it unlikely that variations in the radiation field
alone could be responsible.  Even with many optimizing assumptions, we
find that the effect on $R$ is not likely to be observable in more
than two lines simultaneously.  Our simulations show that either the
line will already be depopulated -- and $ \left<R\right>$ so depressed
as to be itself undetectable, let alone variations in $R$ -- or the
line will be mostly unaffected, and any variation due to stellar
pulsations will be too small to be observable.

We do note that even where our simulations found potentially
observable variation, it would likely be more challenging to measure
in practice than in our model.  As mentioned above, our
phenomenological model has been tailored to produce the maximum
detectable effect, and any attempt to include more ingredients will
reduce even the effects seen here.  In hotter O~stars, line formation
is more complex, with the possibility of photo-absorption of X-rays in
the wind, diluting the impact of changes in the radiation field such
as those modeled here.  In a non-radial pulsator, temperature and
radius do not vary smoothly across the stellar surface, minimizing the
spectral variations and the changes in dilution at the location of the
gas.  Further, adopting a multiple-temperature gas with varying
ionization fractions for each element may decrease the response of the
level populations in the gas and thus of $R$ in each ion.

Finally, a stellar spectrum is not actually a blackbody.  In any
region of the spectrum that is generating UV photons in the
appropriate range, there will be photospheric absorption lines.  If an
absorption line corresponds precisely to the pumping frequency, $R$
will not be affected, either by the UV flux or by its variation. If
the X-ray emitting gas is spatially distributed within the wind, it is
likely that an absorption line will fall somewhere within its range of
Doppler-shifted velocities.  Further, the Lyman edge means that there
will be fewer photons available below 912${\rm \AA}$ than are present
in our model. In any atom with He-like ion pumping wavelength below
the Lyman edge, \eg~silicon and all heavier elements, the true impact
on $R$ will be less than that in our model and $R$ will be closer to
$R_0$.
 
 For these reasons, it seems likely that changes in the radiation
field will not be able to fully account for the reported time
variations in $R$ in hot stars across several lines. Instead, future
studies must explore the impact of changes in density surrounding the
star, through mass-loss changes, clumping in the stellar winds,
ejection events and magnetic field effects.  Such models will provide
us with new tools for understanding the environment of these hot
stars, and their interactions with it.

\begin{acknowledgements} The authors are grateful to Joy Nichols and
Wayne Waldron for drawing our attention to the interesting problem of
variable X-ray line ratios.  We are also indebted to the referee,
Maurice Leutenegger, for conversations that greatly improved the
paper.  Ignace is indebted to David P. Huenemoerder for his assistance
with the use of the ISIS tool \citep{houck2000} for accessing basic
atomic data and CIE models.  This research was supported by NSF grant
AST-0922981, NASA ATFP award NNH09CF39C, and NASA {\em XMM-Newton}
award NNX09AP48G.
\end{acknowledgements}


\bibliographystyle{aa}     

\end{document}